\documentclass[namedreferences]{SolarPhysics}
\pdfoutput=1
\usepackage[optionalrh]{spr-sola-addons} % For Solar Physics 

\usepackage{graphicx}        % For eps figures, newer & more powerfull
\usepackage{amssymb}        % useful mathematical symbols  
\usepackage{upgreek}            
\newcommand{\etal}{{\it et al.}}

\newcommand{\curl}{ {\bf \nabla} \times}

\newcommand{\apj}{    {\it Astrophys. J.}}

\newcommand{\mnras}{  {\it Mon. Not. Roy. Astron. Soc.}}

\newcommand{\pasa}{   {\it Pub. Astron. Soc. Aust.}}

\newcommand{\solphys}{{\it Solar Phys.}}

\begin{document}

\begin{article}

\begin{opening}

\title{Time-Distance Modelling In A Simulated Sunspot Atmosphere}

\author{H.~\surname{Moradi}$^{1}$\sep
        P.S.~\surname{Cally}$^{1}$
	}
\runningauthor{H. Moradi, P.S.Cally}
\runningtitle{Time-Distance Modelling in a Simulated Sunspot Atmosphere}

\institute{$^{1}$ Centre for Stellar and Planetary Astrophysics, School of Mathematical Sciences, Monash University, Victoria 3800, Australia \email{hamed.moradi@sci.monash.edu.au}}
               
\date{Received: 11 March 2008 / Accepted: 14 April 2008}     
             
\begin{abstract}
In time-distance helioseismology, wave travel times are measured from the cross-correlation between Doppler velocities recorded at any two locations on the solar surface. However, one of the main uncertainties associated with such measurements is how to interpret observations made in regions of strong magnetic field. Isolating the effects of the magnetic field from thermal or sound-speed perturbations has proved to be quite complex and has yet to yield reliable results when extracting travel times from the cross-correlation function. One possible way to decouple these effects is by using a 3D sunspot model based on observed surface magnetic-field profiles, with a surrounding stratified, quiet-Sun atmosphere to model the magneto-acoustic ray propagation, and analyze the resulting ray travel-time perturbations that will directly account for wave-speed variations produced by the magnetic field. These artificial travel-time perturbation profiles provide us with several related but distinct observations: \textit{i}) that strong surface magnetic fields have a dual effect on helioseismic rays -- increasing their skip distance while at the same time speeding them up considerably compared to their quiet-Sun counterparts, \textit{ii}) there is a clear and significant frequency dependence of both skip-distance and travel-time perturbations across the simulated sunspot radius, \textit{iii}) the negative sign and magnitude of these perturbations appears to be directly related to the sunspot magnetic-field strength and inclination, \textit{iv}) by ``switching off'' the magnetic field inside the sunspot, we are able to completely isolate the thermal component of the travel-time perturbations observed, which is seen to be both opposite in sign and much smaller in magnitude than those measured when the magnetic field is present. These results tend to suggest that purely thermal perturbations are unlikely to be the main effect seen in travel times through sunspots and that strong, near-surface magnetic fields may be directly and significantly altering the magnitude and lateral extent of sound-speed inversions of sunspots made by time-distance helioseismology. 
\end{abstract}
\keywords{Helioseismology, Direct Modelling; Sunspots, Magnetic Fields; Magnetic Fields, Models}
\end{opening}
%-------------------------------------------------

\section{Introduction}
Time-distance helioseismology is a powerful diagnostic tool used in local helioseismology to probe the subsurface structure and dynamics of the solar interior, in particular in and around solar active regions. To date however, results obtained by time-distance helioseismology have not directly accounted for the effects of the magnetic field on the wave-speed in travel-time perturbation maps, forward modelling or inversions, but have indirectly included magnetic effects only through their influence on the acoustic properties of the medium (\textit{e.g.} the sound speed). Standard forward-modelling is based on a number of assumptions including, but not limited to, Fermat's Principle and the ray approximation (\textit{e.g.} \opencite{kds00}; \opencite{zkd01}; \opencite{hrt05}), the Fresnel-Zone approximation (\textit{e.g.} \opencite{jetal01}; \opencite{sebetal04}) and the Born approximation (\textit{e.g.} \opencite{cbk06}). These models do not include any provision for surface effects. In fact, no standard local-helioseismic method includes provisions for contributions from near-surface magnetic fields. 

Recent work in sunspot seismology has pointed to the significant influence of near-surface magnetic fields and possible contamination due to their effects in helioseismic inversions for sound speed beneath sunspots \cite{sebraj}. Prior to this, a number of other very important results have highlighted the complications of interpreting helioseismic observations (in particular, the interaction of $p$ modes) in the near-surface regions of sunspots (see \textit{e.g.} \opencite{fan95}; \opencite{ccb03}; \opencite{lb2005}; \opencite{han2005}; \opencite{sc2006}; \opencite{bb2006}).

The key issues are \textit{i}) how to successfully model the effects of wave-speed inhomogeneities thought to be produced by the magnetic field in solar active regions, \textit{ii}) how to isolate such effects from those thought to be associated with temperature, flow perturbations, and other observational constraints and effects, and finally \textit{iii}) how will inferences made about subsurface structure change as a result of incorporating these effects into the modelling process? Efforts to address these issues both observationally and computationally have been largely unsuccessful, mainly because of a general lack of understanding of the process involved. But there is some light at the end of the tunnel, as there are currently under development a number of robust magnetohydrodynamical (MHD) simulations modelling helioseismic data and wave propagation that may aid our understanding considerably in the near future (\textit{e.g.} \opencite{cgd08}; \opencite{hd07}). In this work, we shall attempt to address some of these outstanding issues by using helioseismic ray theory to perform forward modelling of helioseismic rays in a simulated sunspot atmosphere with the aim of modelling the magneto-acoustic ray propagation and analysing the resulting artificial ray travel-time perturbations that will directly account for wave-speed variations produced by the magnetic field. We shall also address the problem of trying to isolate and analyze the thermal contributions to the observed travel-time perturbations using our simulations. 

\section{The Sunspot Model}
The axisymmetric sunspot model chosen for our analysis consists of a non-potential, untwisted, magnetohydrostatic sunspot model constrained to fit observed surface magnetic field profiles. The surface field is therefore quite realistic, which is important because there is evidence \cite{sc2006} that magnetic effects on helioseismology are dominated by the top Mm.   
\begin{figure}[h]
\begin{center} 
\begin{tabular}{cc}
\hspace*{-6mm}
\includegraphics[trim= 4mm 0mm 4mm 0, clip, width=14.5pc]{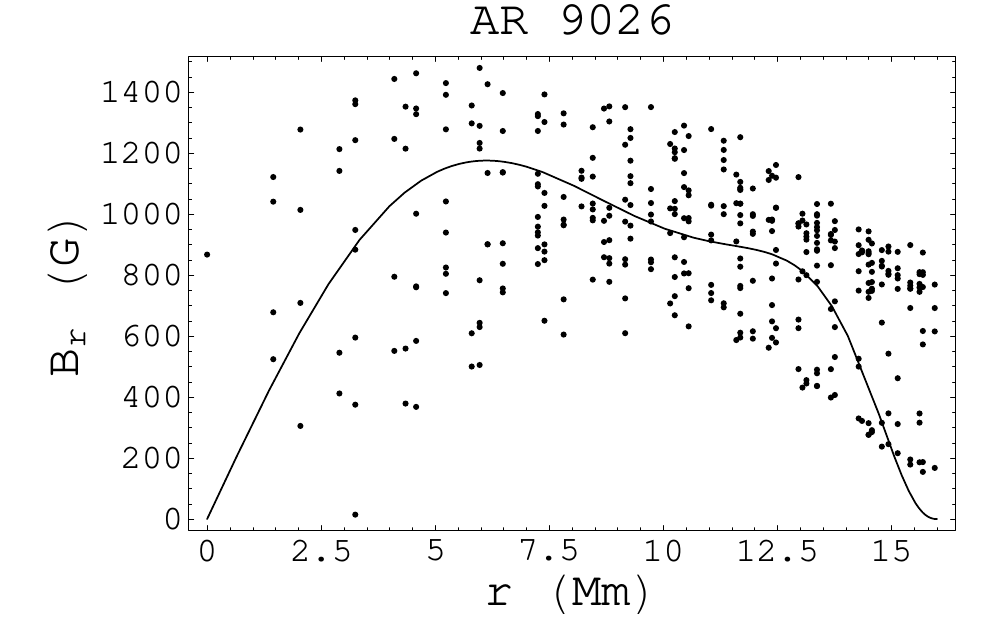}&
\hspace*{-5mm}
\includegraphics[trim= 4mm 0mm 4mm 0, clip, width=14.5pc]{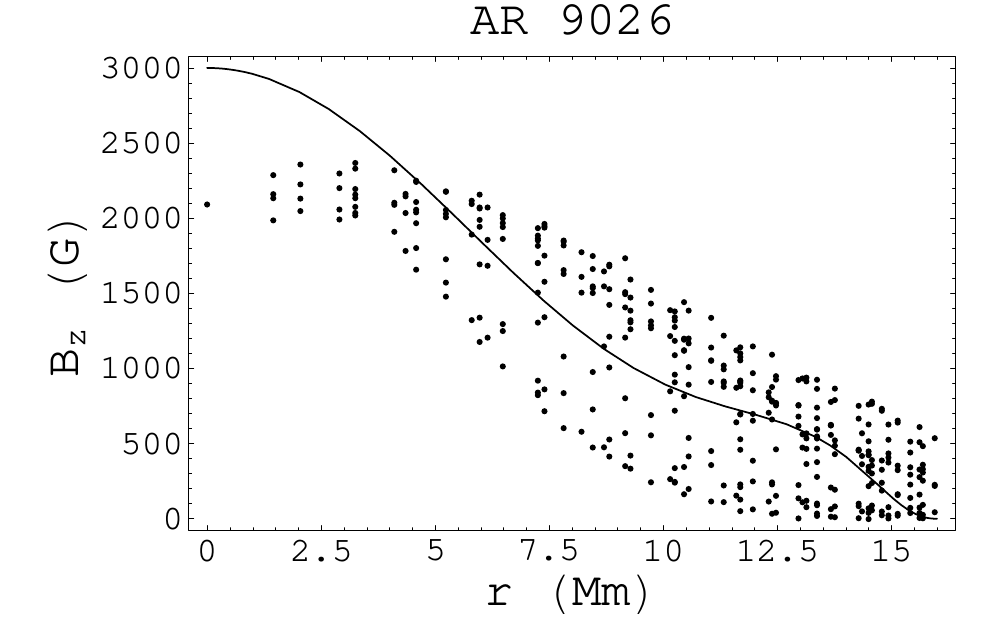}
\end{tabular}
\includegraphics[trim= 4mm 0mm 4mm 0, clip, width=14.5pc]{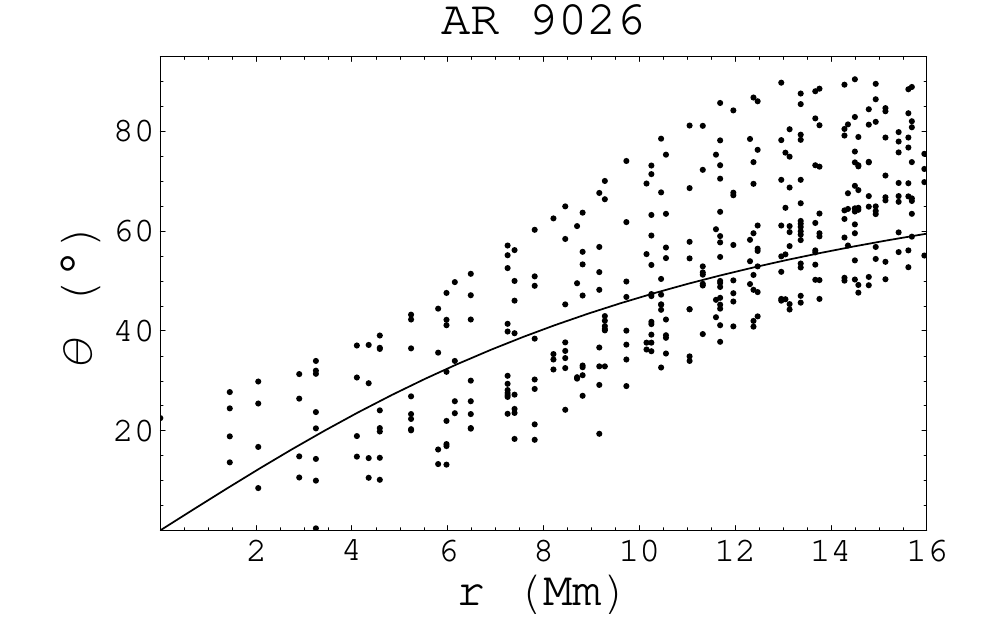}
\end{center}
\caption{Plots of the radial ($B_r$, left), vertical components of the observed magnetic field ($B_z$, right) and magnetic field inclination from the vertical ($\theta^{\circ}$) as derived from IVM surface magnetic field profiles of Active Region (AR) 9026 on 5 June 2000, shown as a function of sunspot radius ($r$, Mm). Solid lines indicate constrained polynomial fits. Values of $B$ are shown in Gauss (G).}
\label{fig:brbz}
\end{figure}

The sunspot also needs to be surrounded by an unperturbed, stratified atmosphere. The background model employed consists of a Global Oscillation Network Group (GONG) Model S atmosphere \cite{cd1996} (obtained from the \{L5BI.D.15C.PRES.960126.AARHUS\} Model S package). The preferred surface field configuration of the flux tube was derived from constrained polynomial fits to the observed scatter plots of the radial ($B_r$) and vertical ($B_z$) surface magnetic field profiles (see Figure~\ref{fig:brbz}) of AR 9026 on 5 June 2000 -- a fairly symmetrical sunspot near disk-centre, ideal for helioseismic analysis -- obtained from IVM (Imaging Vector Magnetograph) vector magnetograms (see \inlinecite{mick96} for more details regarding the observations). We note that in the $B_z$ profile of AR 9026, the vertical-field strength tends to \textit{decrease} to around 2~kG as it approaches $r=0$. We find this highly improbable for a sunspot, so we extrapolate to a peak field of 3~kG for our model at $r=0$. (A separate analysis was conducted for the model with the (unrealistic) peak field of 2~kG. As expected, the only difference we observed was the magnitude of the perturbations produced being slightly smaller than the ones we report in Section 4. All other results appeared to be identical). The fits of $B_r$ and $B_z$ are then used to derive an analytical form for the potential function, 
\begin{equation}
\Psi(r,z)=\psi_0 \left(\frac{R_0 r}{r_b(z)}\right)
\end{equation}
where $\psi_{0}$ is the derived surface field at the surface ($z=Z_0$), the radius of the sunspot at the surface ($r=R_0$) is fixed at $R_0=16$~Mm. Instead of a current sheet along the boundary, we prescribe an analytical form for the outermost field line, 
\begin{equation}
r_b(z)=\frac{R_0-R_m}{(1-c)\mathrm{e}^{-(z-Z_0)/\lambda}+c}+R_{m},
\end{equation} 
where the field strength drops to zero and $R_m$ and $c$ are free parametres. We ensure that all calculations (\textit{e.g.} change in pressure, density, \textit{etc.}) made across the boundary layer/transition region between the sunspot atmosphere and the external environment are both consistent and continuous along $r_b$.  
%\begin{figure}[h]
%\begin{center} 
%\begin{tabular}{cc}
%\hspace*{-6mm}
%\includegraphics[trim= 14mm 9mm 14mm 9mm, clip, width=15pc]{dopp9026b}&
%\hspace*{-10mm}
%\includegraphics[trim= 14mm 9mm 14mm 9mm, clip, width=15pc]{cont9026b}&
%\end{tabular}
%\includegraphics[trim= 14mm 9mm 14mm 9mm, clip, width=15pc]{mag9026b}
%\end{center}
%\caption{Plots of the radial ($B_r$, left), vertical components of the observed magnetic field ($B_z$, right) and magnetic field inclination from the vertical ($\theta\degree$) as derived from IVM surface magnetic field profiles of AR 9026 on 5 June 2000, shown as a function of sunspot radius ($r$, Mm). Values of $B$ are shown in Gauss (G).}
%\label{fig:brbz}
%\end{figure}

The next step essentially involves solving the standard equations of magnetohydrostatics (MHS), using the Model S atmosphere and its variables as the quiet-Sun environment. The magnetic pressure and tension resulting from the Lorentz force, 
\begin{equation}
\bf{f}_{L} =\bf{J} \times \bf{B}, 
\end{equation}
are confined within the simulated sunspot atmosphere, where $\mu$ is the magnetic permeability and $\bf{J}= \frac{1}{\mu}(\curl \bf{B})$. 
\begin{figure}
\begin{center}
\includegraphics[width=24pc]{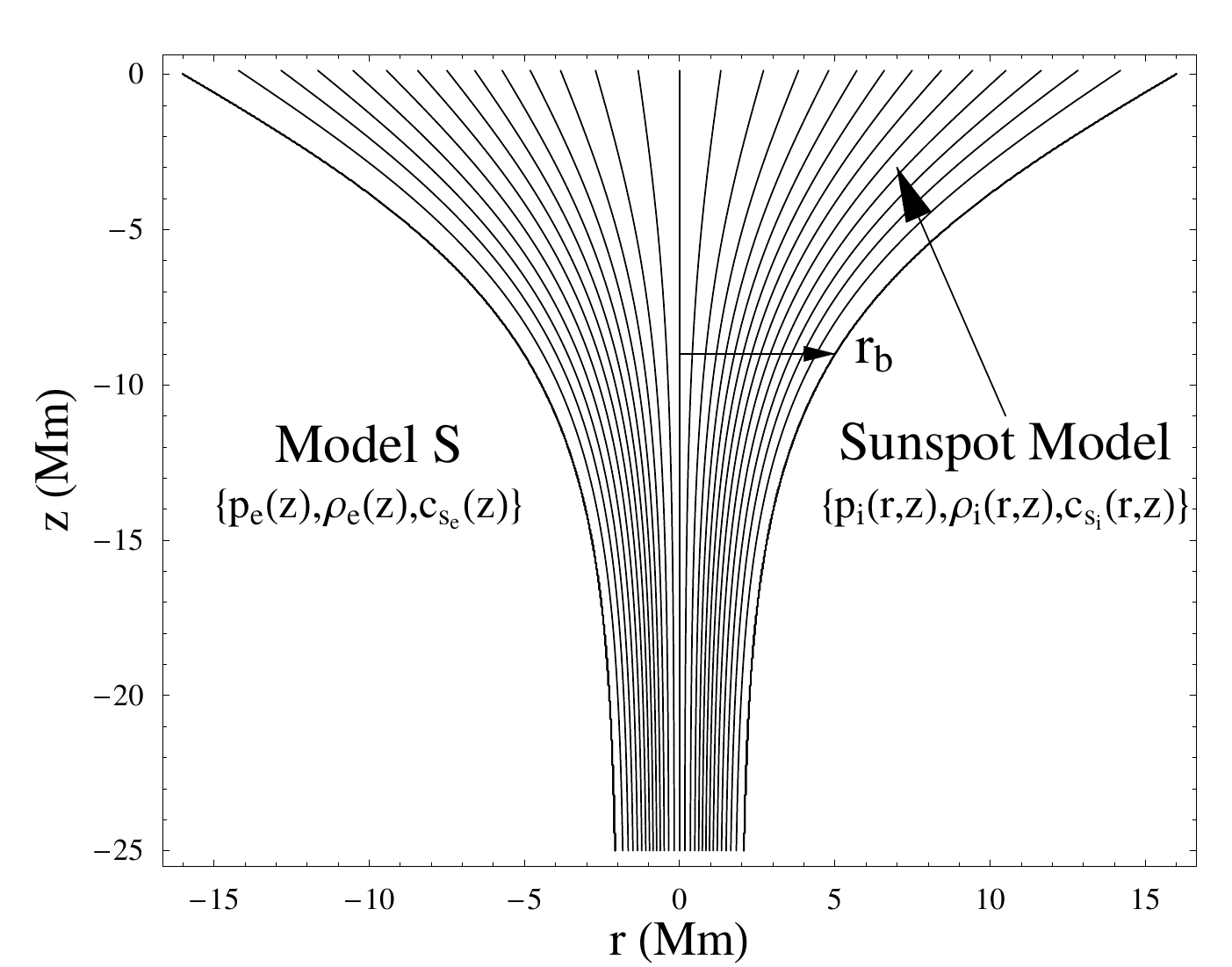}
\end{center}
\caption{The magnetic field configuration for the sunspot model. The field lines plotted indicate equidistant magnetic-flux values. Internal and external (Model S) variables are indicated for reference. $r_b$ represents the radius of the outermost field line, which varies with depth ($z$) along the sunspot radius.}
\label{fig:field}
\end{figure}
The gas pressure $p(r,z)$ is calculated using horizontal force balance, 
\begin{equation}
p_i(r,z)=p_e(z)+\Delta p(r,z)
\end{equation}
where $p_i(r,z)$ and $p_e(z)$ denote internal and external (\textit{i.e.} Model S) pressure respectively and the change in pressure is therefore
\begin{equation}
\Delta p(r,z) = \int^{r}_{r{_b}}f_{L{_r}} \mathrm{d} r 
\end{equation}
which drops to zero as we approach $r_b$. Once the pressure inside the sunspot and along the boundary are known, the density $\rho(r,z)$, can similarly be calculated using vertical force balance,
\begin{equation}
\rho_i(r,z)=\rho_e(z)+\Delta \rho(r,z)
\end{equation}
where the change in density is given by 
\begin{equation}
\Delta \rho(r,z)=\frac{1}{g}\left[f_{L_{z}} - \frac{\partial \Delta p(r,z)}{\partial z}\right]
\end{equation}

This is essentially all that is required to then compute the modified sound speed or thermal profile of the sunspot atmosphere,
\begin{equation}
c^2_{s_{i}}(r,z)=c^2_{s_{e}}(z)+\Gamma_1(z)\left[\frac{p_i(r,z)}{\rho_i(r,z)} - \frac{p_e(z)}{\rho_e(z)} \right],
\end{equation} 
while for the sake of simplicity, assuming the ratio of specific heat ($\Gamma_1$) that appears in the sound speed is the same function of height as it is in the external atmosphere. Finally, all that is left is to calculate the Alfv\'en speed, 
\begin{equation}
a^2(r,z) =\frac{1}{\mu\rho_i(r,z)}[B^2_r + B^2_z].
\end{equation}   

\begin{figure}[h]
\begin{center}
\begin{tabular}{cc}
\hspace*{-10mm}
\includegraphics[width=15pc]{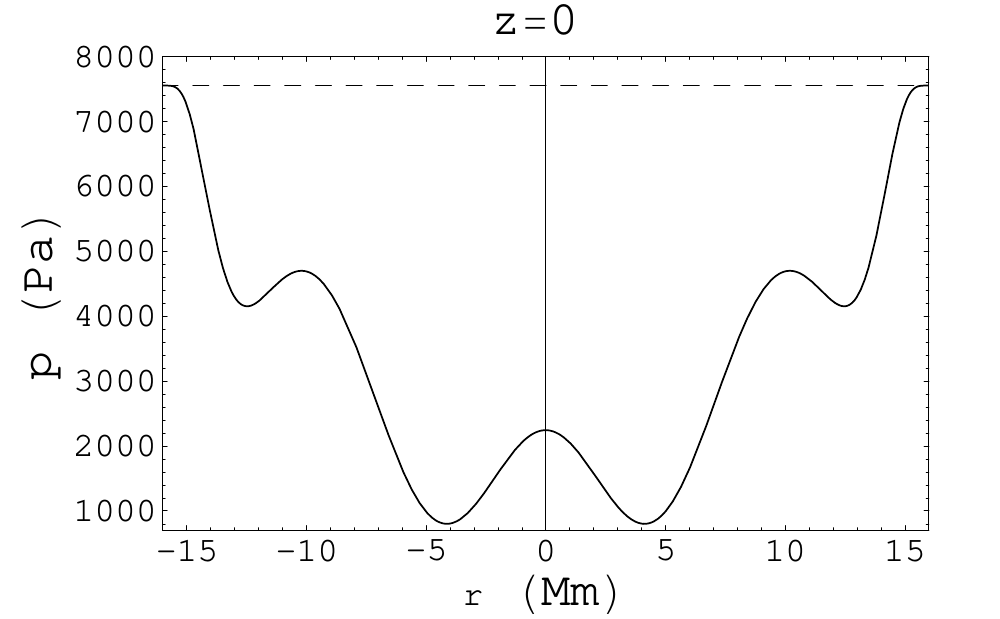}&
\hspace*{-7mm}
\includegraphics[width=15pc]{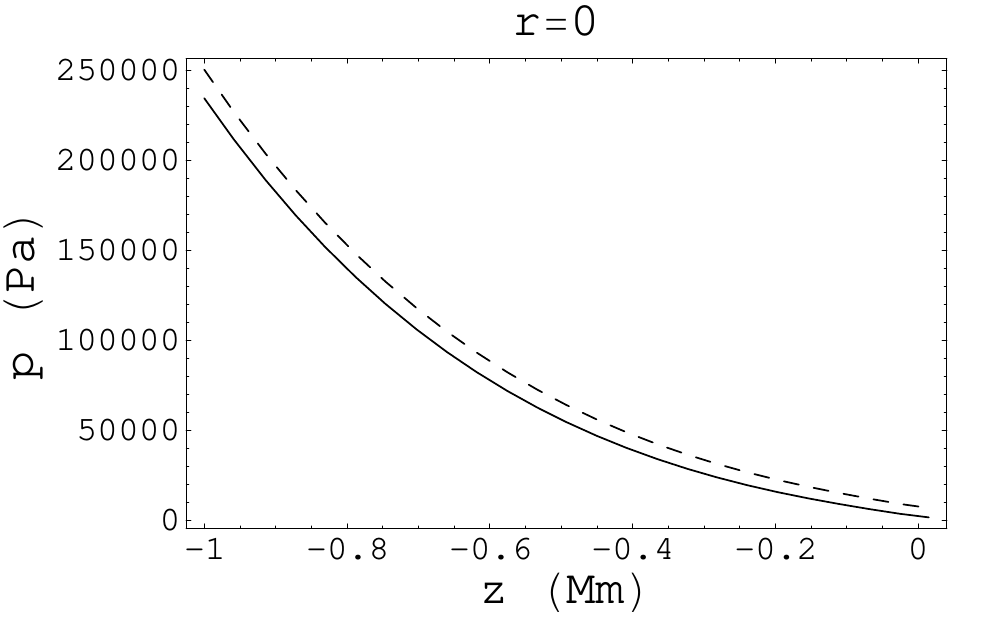}\\
\hspace*{-10mm} 
\includegraphics[width=15pc]{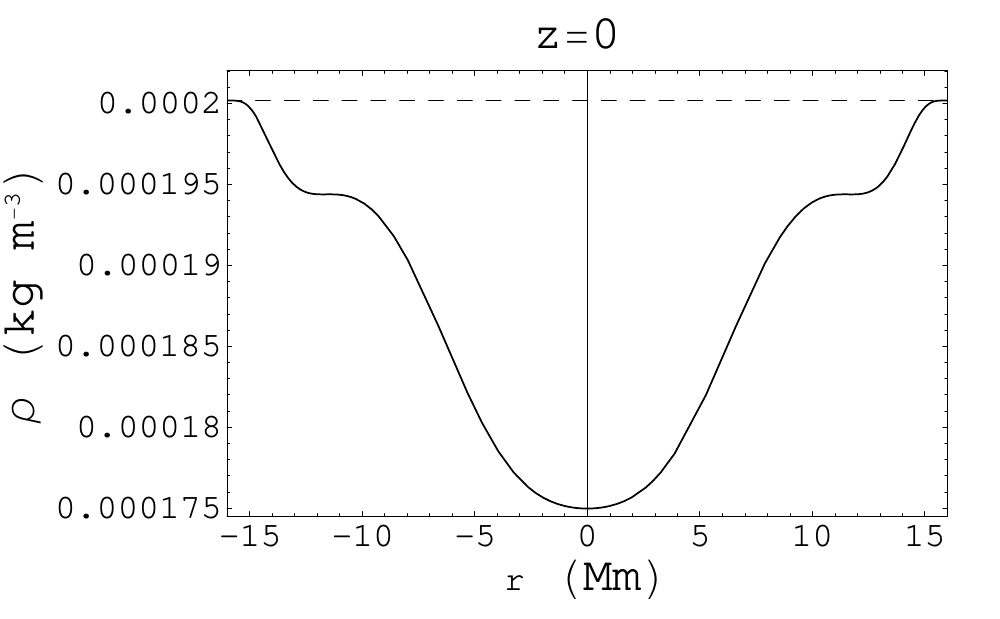}&
\hspace*{-7mm}
\includegraphics[width=15pc]{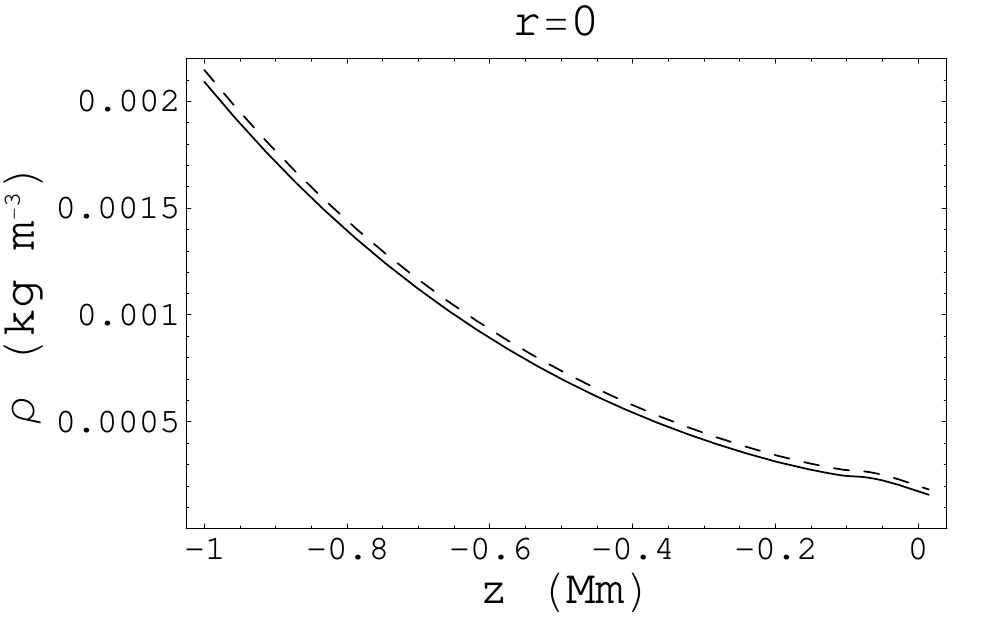}\\
\hspace*{-10mm}
\includegraphics[width=15pc]{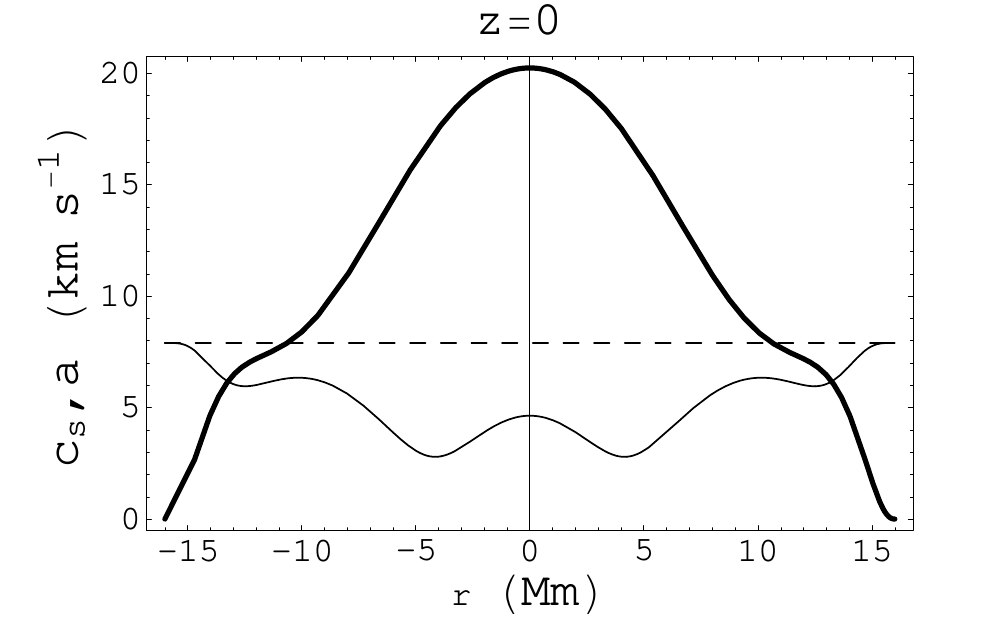}&
\hspace*{-7mm}
\includegraphics[width=15pc]{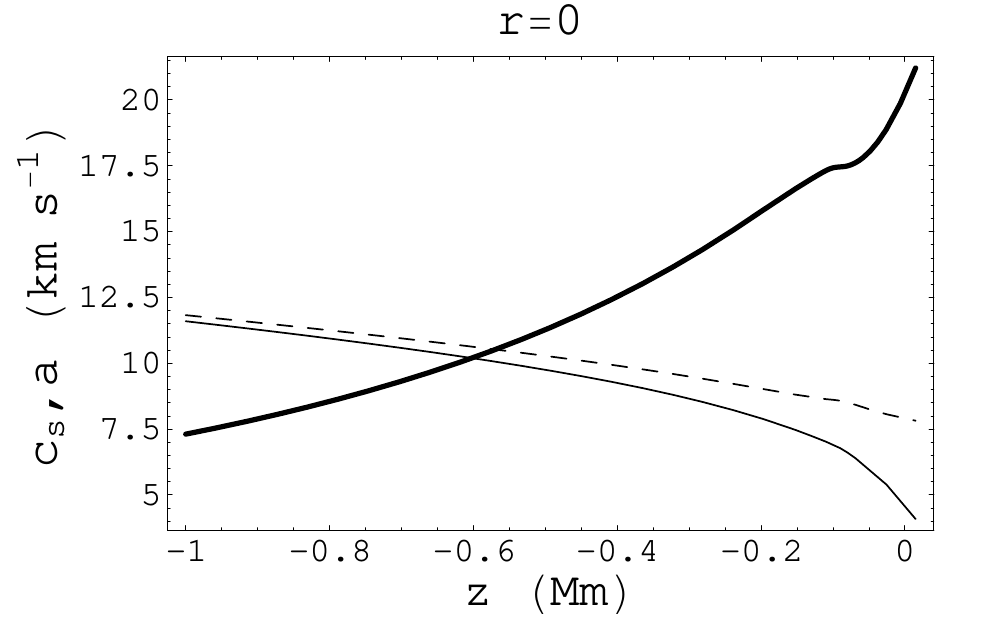}
\end{tabular}
\end{center}
\caption{Internal pressure ($p$), density ($\rho$), sound ($c_s$), and Alfv\'en ($a$) speed profiles of the sunspot model with an external GONG Model S atmosphere. Left-hand coloumn profiles are calculated along the surface of the sunspot ($z=0$), while right-hand coloumn profiles are calculated along the axis of the sunspot ($r=0$). Internal profiles are indicated by solid lines in all plots. The thick solid line in the bottom two panels indicate Alfv\'en speeds. The dashed lines represent GONG Model S values in all plots. }
\label{fig:models}
\end{figure}

\begin{figure}[h]
\begin{center}
\hspace*{-5mm}
\includegraphics[trim= 6mm 1mm 6mm 2mm, clip, width=24pc]{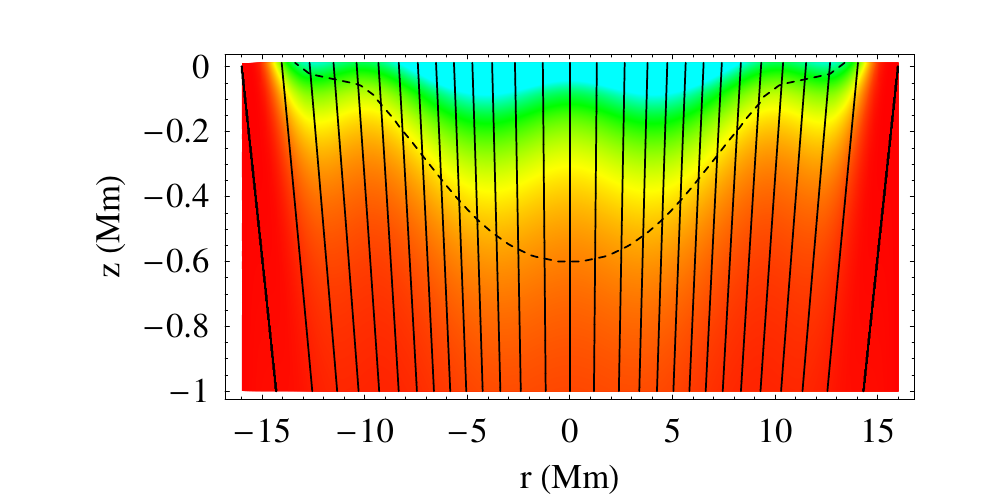}
\end{center}
\caption{The thermal profile ($c^2_s$) in the top 1~Mm of the sunspot. Lighter coloured contours (\textit{i.e.} cyan/green) indicate regions of decreased sound speed (cooler regions) under the sunspot surface, while darker (hotter) regions (\textit{i.e.} orange/red) are indicative of areas of enhanced sound speed. The dashed line marks the position of the $a=c_s$ layer. Field lines are over-plotted}
\label{fig:thermals}
\end{figure}

Some of the important internal properties of the resulting sunspot model (\textit{e.g.} pressure, density, sound and Alfv\'en speeds) are shown in Figure~\ref{fig:models}. The external (Model S) profiles for each variable are also shown for reference. The near-surface thermal structure of the sunspot and the ($a=c_s$) equipartition depth is also shown for reference in Figure~\ref{fig:thermals}. We can clearly see the modified sound-speed structure ($c^2_s$) as a result of the magnetic field in this image. It is interesting to note that in our (simple) model the region of decreased sound-speed does not appear to extend as deep as 3D time-distance inversions of the real Sun have suggested. Estimates for the lateral extent of the decreased sound-speed region using tomographic imaging of the sub-surface layers of sunspots have ranged from depths of approximately $z=-2.4$ to $z=-3.5$~Mm using the Born and ray approximations respectively \cite{cbk06}. Nevertheless, the sunspot model exhibits the broad features expected of a real sunspot, and presents a useful test case.    

\section{Ray Path Calculations}
\label{sec:raypaths}
\begin{figure}[t]
\begin{center}
\includegraphics[trim=0mm 18mm 0mm 18mm, clip, width=22pc]{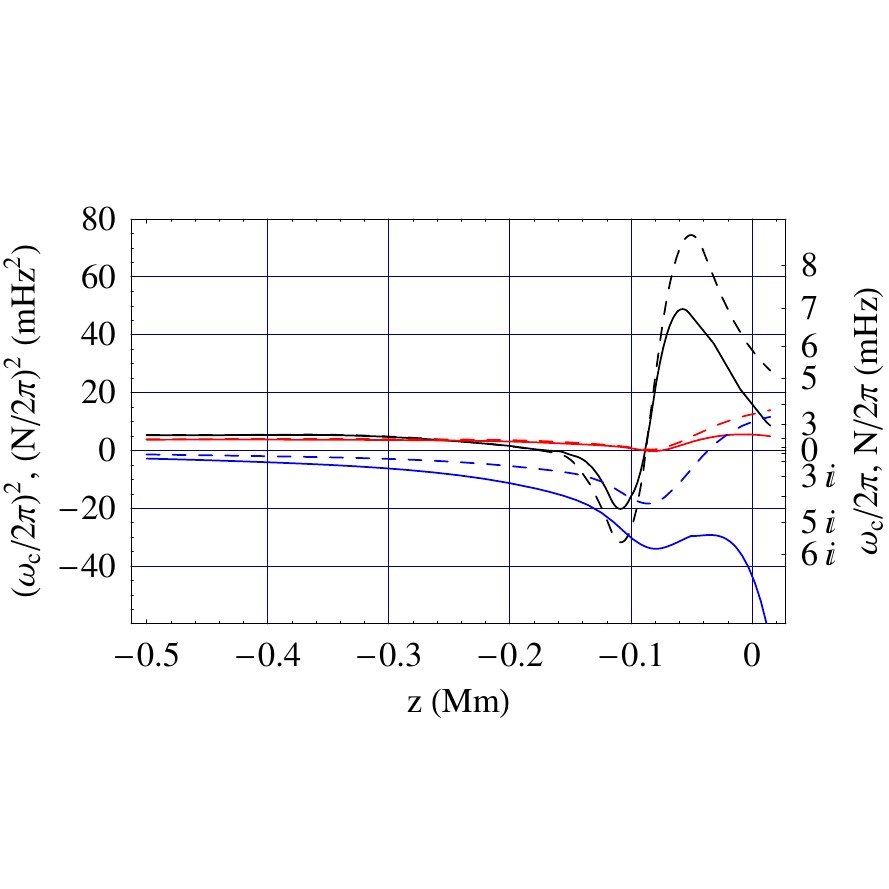}
\end{center}
\caption{Plots of the various forms of the acoustic cutoff ($\omega_c$) and Brunt-V\"ais\"al\"a ($N$) frequencies. The later is indicated by a blue, solid line inside the sunspot atmosphere and dashed blue line indicating Model S values. The solid black line indicates the acoustic cutoff frequency $\omega_c$ for the sunspot atmosphere, while the dashed black line indicates Model S values. The isothermal form, $\omega_{c{_i}}$ is indicated by the solid red line for the sunspot atmosphere, dashed red line indicates Model S values.}
\label{fig:omegac}
\end{figure}

The ray paths are calculated in Cartesian geometry, in the realm of frequency dependent ray paths described by \inlinecite{bc}, with the complete form of the three-dimensional dispersion relation:
\begin{eqnarray}
%\begin{split}
\mathcal{D} = \omega^2 \omega^2_c a^2_y k^2_h + (\omega^2 - a^2 k^2_\parallel) \times [\omega^4 - (a^2+c^2) \omega^2k^2 \nonumber \\ + a^2c^2k^2k^2_\parallel + c^2N^2k^2_h - (\omega^2-a^2_zk^2)\omega^2_c] = 0,
%\end{split}
\label{eq:disp}
\end{eqnarray}
where $k_h$ and $k_\parallel$ are the horizontal and parallel components of the wave-vector $\bf{k}$ and
\begin{equation}
N^2=\frac{g}{H_\rho}-\frac{g^2}{c^2}
\end{equation}
is the squared Brunt-V\"ais\"al\"a  frequency, with $g$ being the gravitational acceleration, $H_\rho (z)$ the density scale height, and $H'_\rho=\mathrm{d} H_\rho / \mathrm{d} z$ and $\omega^2_c$ is the square of the acoustic-cutoff frequency. For completeness, we calculate the raypaths using two forms of $\omega_c$. The most commonly used form
\begin{equation}\label{eq:wc}
\omega^2_c = \frac{c^2}{4H^2_\rho}(1-2H'_\rho),
\label{eq:gough}
\end{equation}
exhibits an extended sharp spike around $z=-100$~km (see Figure~\ref{fig:omegac}). This form of $\omega_c$ is often used by helioseismologists. However, as \inlinecite{cally06} points out, this sharp spike in the cutoff frequency is inconsistent with the WKB assumption of slowly varying coefficients on which $\mathcal{D}$ is based. A much smoother isothermal form,
\begin{equation}
\omega_{c_{i}}=c/2H,
\label{eq:iso}
\end{equation}
is consistent with the derivation of $\mathcal{D}$, and does not suffer from the spike (see Figure~\ref{fig:omegac}). Unless otherwise stated, all results shown here utilize $\omega_{c_{i}}$. (Simulations using the form of $\omega_c$ in Equation~(12) were also conducted, the results being very similar to those reported in Section 4, expect for a certain amount of unsmoothness being present in the travel-time perturbation profiles (mainly affecting shallow rays which are more sensitive to the reflecting boundary near the surface) as a result of using the more rigid form of $\omega_c$). Naturally, the magnetic field slightly modifies both $\omega_c$ and $\omega_{c_{i}}$, the results of which can be seen in Figure~\ref{fig:omegac}. 

Following \inlinecite{wein62}, the construction of $\bf{k}$ is completed by specifying the governing equations of the ray paths
\begin{equation}
\frac{\mathrm{d}\mathbf{x}}{\mathrm{d}\tau}=\frac{\partial \mathcal{D}}{\partial \mathbf{k}}
\end{equation}

\begin{equation}
\frac{\mathrm{d}\mathbf{k}}{\mathrm{d}\tau}=-\frac{\partial \mathcal{D}}{\partial \mathbf{x}}
\end{equation}

\begin{equation}
\frac{\mathrm{d}t}{\mathrm{d}\tau}=-\frac{\partial \mathcal{D}}{\partial \omega}
\label{eq:tau}
\end{equation}

\begin{equation}
\frac{\mathrm{d}\omega}{\mathrm{d}\tau}=-\frac{\partial \mathcal{D}}{\partial \tau}
\end{equation}
where $\tau$ parameterizes the progress of a disturbance along the ray path. For a time-independent medium, for which $\partial\mathcal{D}/\partial t=0$ and $\omega$ is constant, the phase function $S(\mathbf{x})$ evolves according to 
\begin{equation}
\frac{\mathrm{d} S}{\mathrm{d}t}=\mathbf{k}\mathbf{\cdot} \frac{\mathrm{d} \mathbf{x}}{\mathrm{d} t}-\omega.
\end{equation}
Hence,
\begin{equation}
\label{eq:phase}
S(\mathbf{x})=\int{\mathbf{k}\mathbf{\cdot}\mathrm{d} \mathbf{x}}-\omega t,
\end{equation}
where the first term (integral) represents the contribution to the phase due to motion along the ray path, and the second term represents the Eulerian part. Since we are only going to be concerned about the change in phase due to motion along the ray path, we can essentially ignore the Eulerian part for the rest of our analysis. 
 
We iteratively find the initial wave-vector ($\mathbf{k}_{\mathrm{init}}$) by using an initial guess which comes from solving $\mathcal{D}=0$ for the wavenumber, assuming the wavevector is in the directions $\alpha$, $\beta$ -- where $\alpha$ and $\beta$ are angles from the vertical and the $x$--$z$ plane respectively of the initial shot. Initially, we initiated the rays from the top of the ray path, adjusting the initial shooting angle ($\alpha$) to obtain the desired range of skip distances. However, given the very sensitive nature of the near-surface region of the sunspot atmosphere, we used a much finer computational grid in the top 1.5~Mm. As a result, we encountered many instances of rays initiated inside evanescent regions (which should obviously be avoided) and also obtaining very shallow rays with little or no helioseismic value. So in order to reduce computation time and also have greater flexibility in choosing the desired range of ray skip distances, we initialized the rays from the minima of their trajectories (essentially the lower turning point of the ray, $z_{\mathrm{bot}}$). Hence, the value of $\alpha$ was fixed at $\alpha=90^{\circ}$, allowing us to adjust the initial shooting depth $z_{\mathrm{bot}}$ to obtain the desired range of skip distances. 

A number of other important points regarding the simulations should also be noted. Firstly, in this paper we only examine the 2D case ($\beta=0$) where rays are confined to the $x$--$z$ plane. Furthermore, by ensuring that the rays remain on the fast-wave branch at all times, we avoid any mode-conversion effects as rays pass through the $a=c_s$ layer (where fast/slow conversion occurs, see Figure~\ref{fig:thermals}). Of course, as numerous works exploring MHD mode conversion in local helioseismology have shown (\textit{e.g.} \opencite{sb92}; \opencite{cb93}; \opencite{cbz94}; \opencite{bc97}; \opencite{cb97}; \opencite{cally00}, \citeyear{cally06}; \opencite{cc03}, \citeyear{cc05}; \opencite{sc2006}), mode transmission and conversion between fast and slow magneto-acoustic waves indeed occurs as rays of helioseismic interest pass through the $a=c_s$ equipartition level and have distinct effects on helioseismic waves that should not be ignored. But in our current analysis (and as with actual time-distance inversions) we do not directly account for these effects. As a result the complexities of the ray-path calculations are greatly reduced. We also note that we ignore any finite-wavelength effects and filtering of observations in our simulations.   

The computational ray propagation grid extends across the 16~Mm radius of the sunspot model in regular 1~Mm spatial increments in the horizontal $x$-direction and down to a depth of 25~Mm in the vertical $z$-direction, employing a much finer grid spacing in the top 1.5~Mm, followed by 1~Mm increments down to a depth of 25~Mm. The cutoff height (depth) for all rays propagated in the grid was fixed at $z=-0.1$~Mm, regardless of frequency. This computational grid, though not exhaustive, allows us to obtain the desired range of skip distances required to replicate the ``centre-to-annulus'' skip distance geometry (\textit{i.e.} averaging rays from a central point/pixel to a surrounding annulus of different sizes to probe varying depths beneath the solar surface) often employed in time-distance helioseismology for the derivation of mean travel-time perturbation maps (see \inlinecite{gblr} for a more comprehensive description of this process). The 11 standard skip distance bin/annuli ($\Delta$) sizes usually used for these calculations are detailed in Table 1. 
\begin{table}[h]
\caption{The annuli (or skip-distances) geometries used to bin the ray travel-time measurements.}
\begin{tabular}{rr}
\hline
$\Delta$ & Pupil Size (Mm)\\
\hline
1 &   3.7 - 8.7\\
2 &   6.2 - 11.2\\
3 &   8.7 - 14.5\\
4 &   14.5 - 19.4\\
5 &   19.4 - 29.3\\
6 &   26.0 - 35.1\\
7 &   31.8 - 41.7\\
8 &   38.4 - 47.5\\
9 &   44.2 - 54.1\\
10 &  50.8 - 59.9\\
11 &  56.6 - 66.7\\ 			
\hline
\end{tabular}
\label{tab:bins}
\end{table}
\vspace*{-16mm}

\section{Results}
\subsection{Travel-Time and Skip-Distance Perturbations}
The ray propagation grids were computed for three frequencies, $\omega=3.5$, $4$, and $5$~mHz. Both the phase ($t_\mathrm{p}$, associated with the \textit{phase} velocity) and group ($t_\mathrm{g}$, associated with the envelope peak of a wave packet as it travels at the \textit{group} velocity) ray travel times were calculated along each ray path for every radial grid position ($r_\mathrm{spot}$, which is essentially the radial position of the lower turning point of the ray) along the sunspot model. In time-distance helioseismology, centre-to-annulus travel times are extracted from Gaussian wavelet fits -- usually represented by a function of the form
\begin{equation}
W_{\pm}(t)=A\mathrm{e}^{-\gamma^2(t \mp t_\mathrm{g})^2}\cos[\omega_0(t \mp t_\mathrm{p})],
\label{eq:gauss}
\end{equation}
(where all parametres are free) -- to both the positive and negative time parts of the observed cross-correlations \cite{gblr}. However, $t_\mathrm{p}$ is more often used in time-distance literature, primarily as a result of difficulties (mainly observational noise) associated with fitting to the envelope peak. Furthermore, because $t_\mathrm{p}$ is much more independent of the shape of the wave packet than $t_\mathrm{g}$ (as the shape of the wavepacket depends on (unmodelled) mode conversion), we shall also limit our analysis to $t_\mathrm{p}$ calculations in this paper. We identify the phase travel time as
\begin{equation}
t_\mathrm{p}=\frac{S(\mathbf{x})}{\omega},
\end{equation}
which is consistent with the form of $t_\mathrm{p}$ described by the Gaussian wavelet. These travel times are then subtracted from similar ray travel times calculated using the quiet-Sun atmosphere to produce \textit{travel-time perturbation} ($\updelta\tau_\mathrm{p}$) profiles. In general, travel-time \textit{differences} are sensitive to sub-surface flows, while \textit{mean} travel times are sensitive to wave-speed perturbations. However, as our model does not contain flows, we do not need to distinguish directions along ray paths. 
\begin{figure}[h]
\begin{center}
\begin{tabular}{cc}
\hspace*{-3mm}
\includegraphics[trim= 0mm 14mm 0mm 14mm, clip, width=13.5pc]{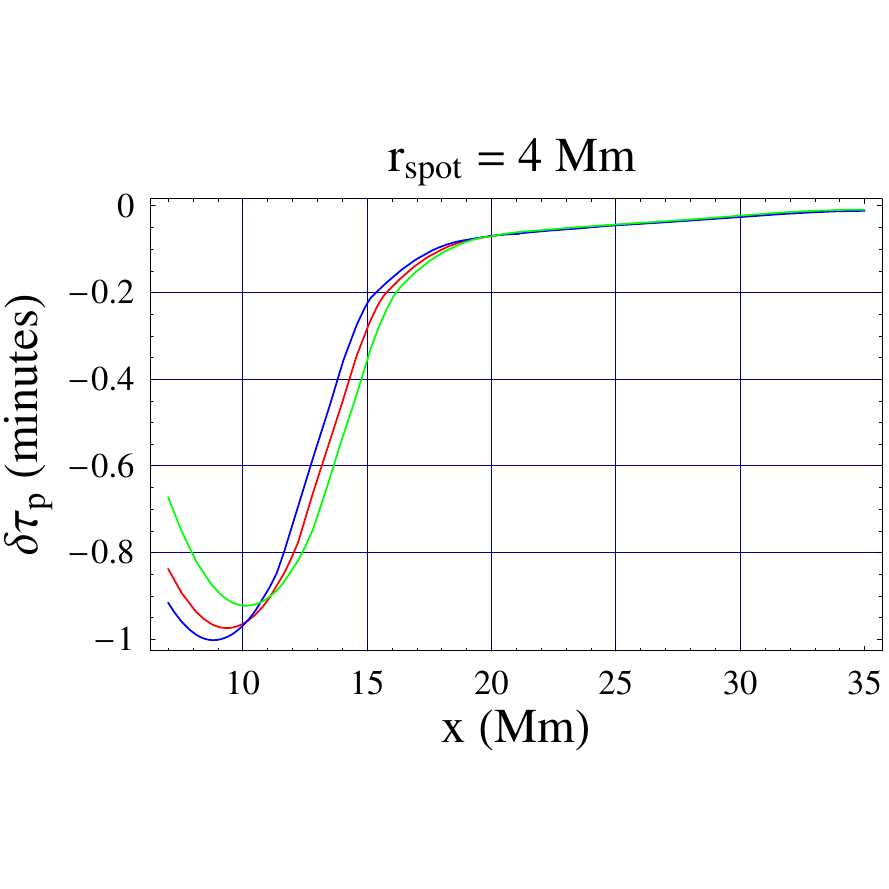}&
\hspace*{-3mm}
\includegraphics[trim= 0mm 14mm 0mm 14mm, clip, width=13.5pc]{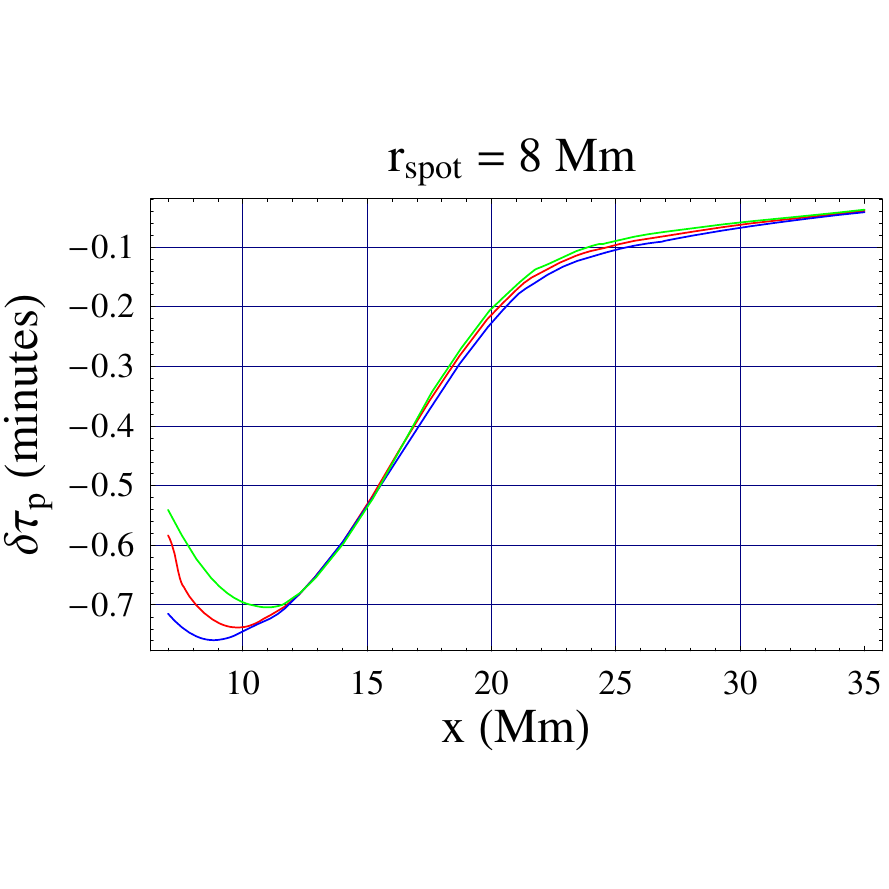}\\
\hspace*{-3mm}
\includegraphics[trim= 0mm 14mm 0mm 14mm, clip, width=13.5pc]{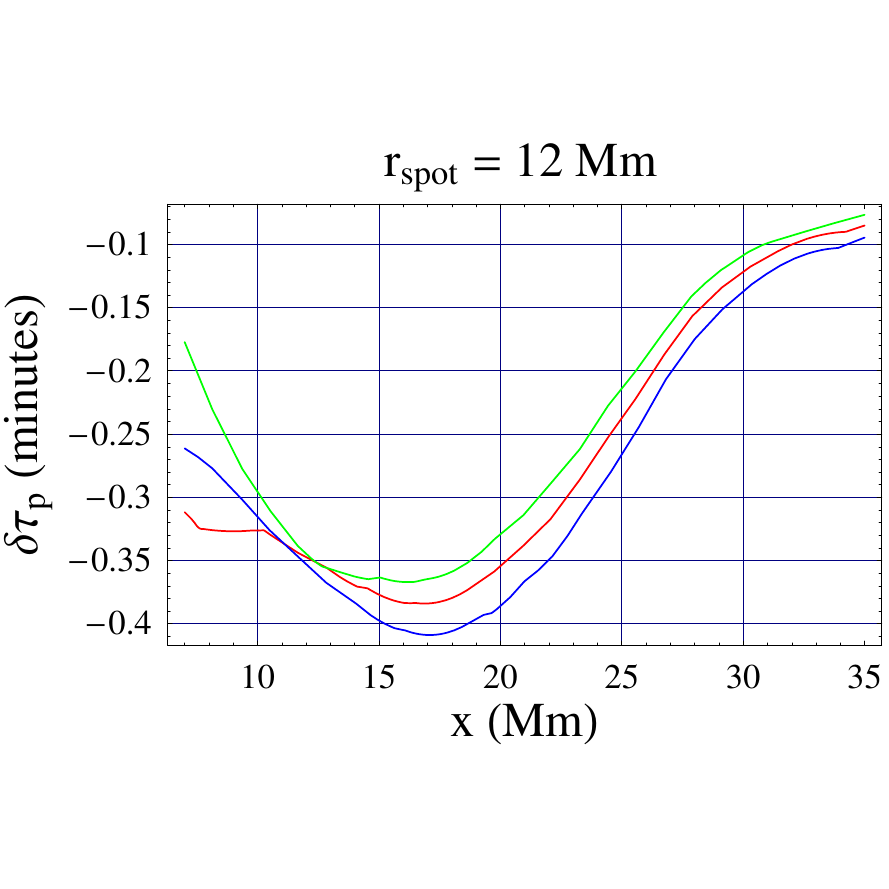}&
\hspace*{-3mm}
\includegraphics[trim= 0mm 14mm 0mm 14mm, clip, width=13.5pc]{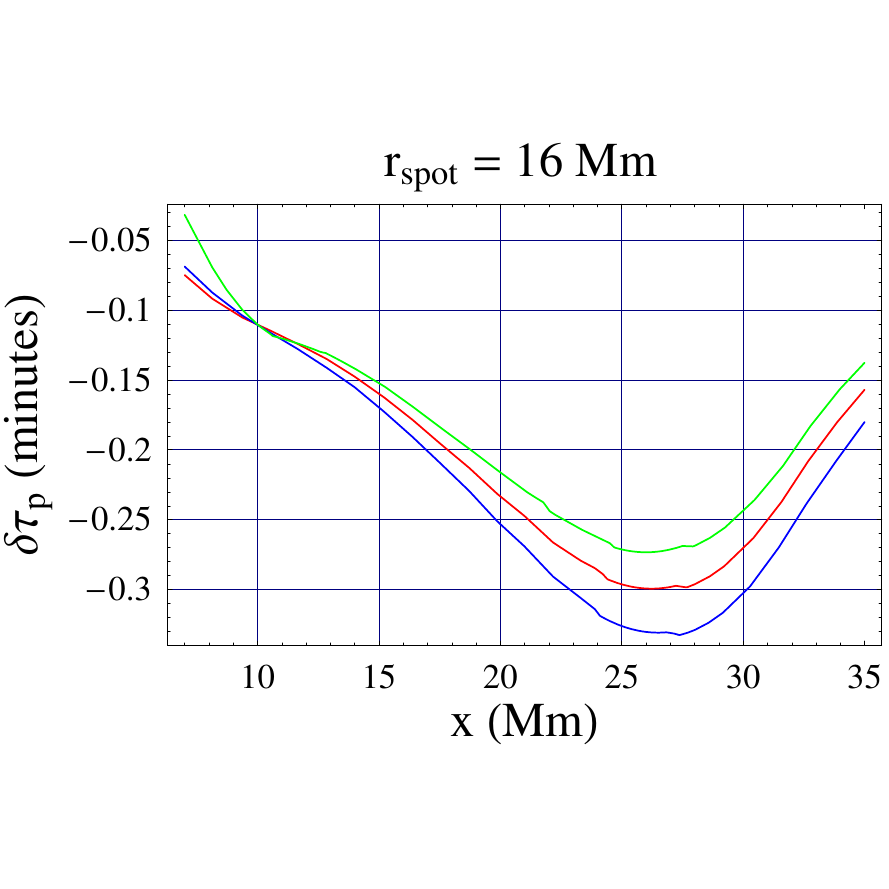}
\end{tabular}
\end{center}
\caption{Travel-time perturbations ($\updelta\tau_\mathrm{p}$) as a function of skip distance ($x$) for $r_{\mathrm{spot}}=4, 8, 12$, and $16$~Mm on the sunspot (where $r_{\mathrm{spot}}$ is the radial position of the lower turning point of the ray), as calculated for three frequencies: $\omega=3.5$ (green), $\omega=4$ (red) and $\omega=5$~mHz (blue).}
\label{fig:deltat}
\end{figure}

In Figure~\ref{fig:deltat} we see some sample $\updelta\tau_\mathrm{p}$ profiles for $r_{\mathrm{spot}}=4, 8, 12$, and $16$~Mm are shown as a function of ray skip distance ($x$) for $\omega=3.5$ (green), 4 (red), and 5 mHz (blue). By and large, there are significant perturbations as we approach the centre of the sunspot (\textit{i.e.} regions associated with stronger surface magnetic field strength). The sign of the perturbations appears to remain exclusively negative, regardless of position on the sunspot. This means that all rays propagated within the simulated sunspot atmosphere are significantly sped up when compared to their Model S counterparts. 

\begin{figure}[h]
\begin{center}
\begin{tabular}{cc}
\hspace*{-3mm}
\includegraphics[trim= 0mm 12mm 0mm 12mm, clip, width=13.5pc]{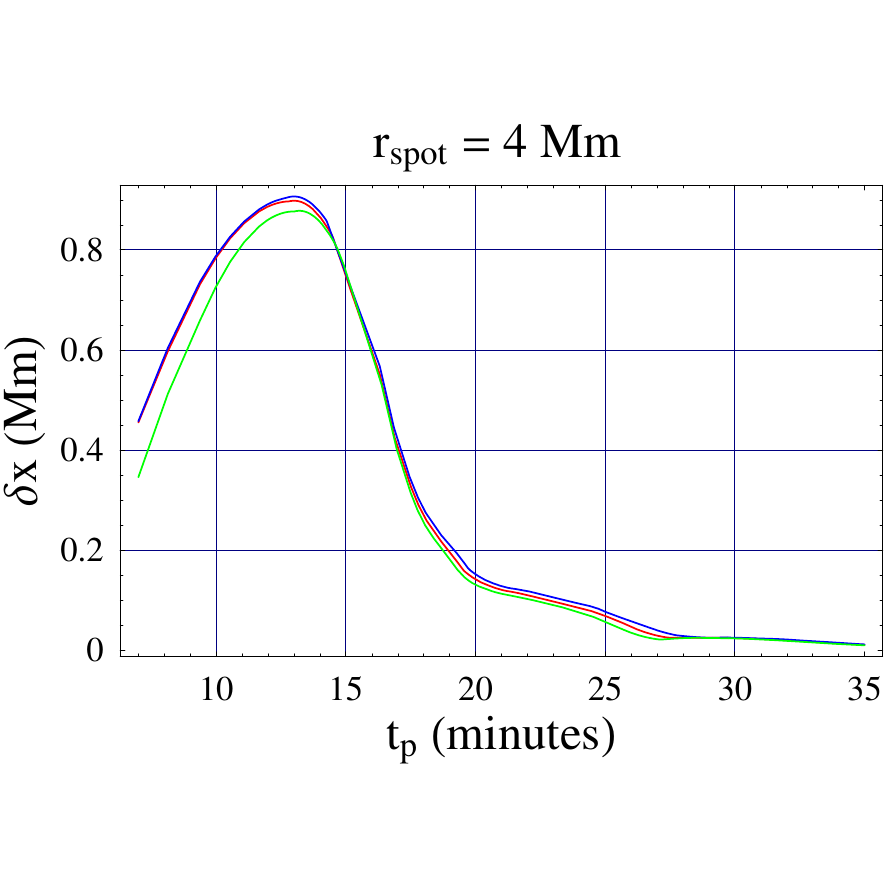}&
\hspace*{-3mm}
\includegraphics[trim= 0mm 12mm 0mm 12mm, clip, width=13.5pc]{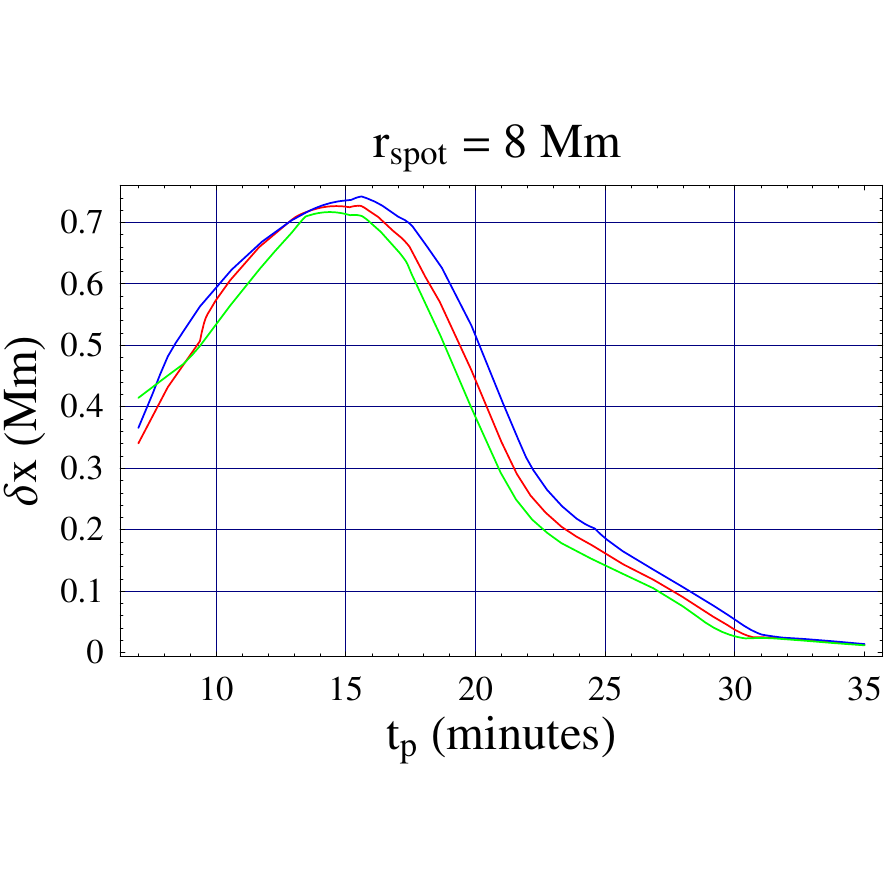}\\
\hspace*{-3mm}
\includegraphics[trim= 0mm 12mm 0mm 12mm, clip, width=13.5pc]{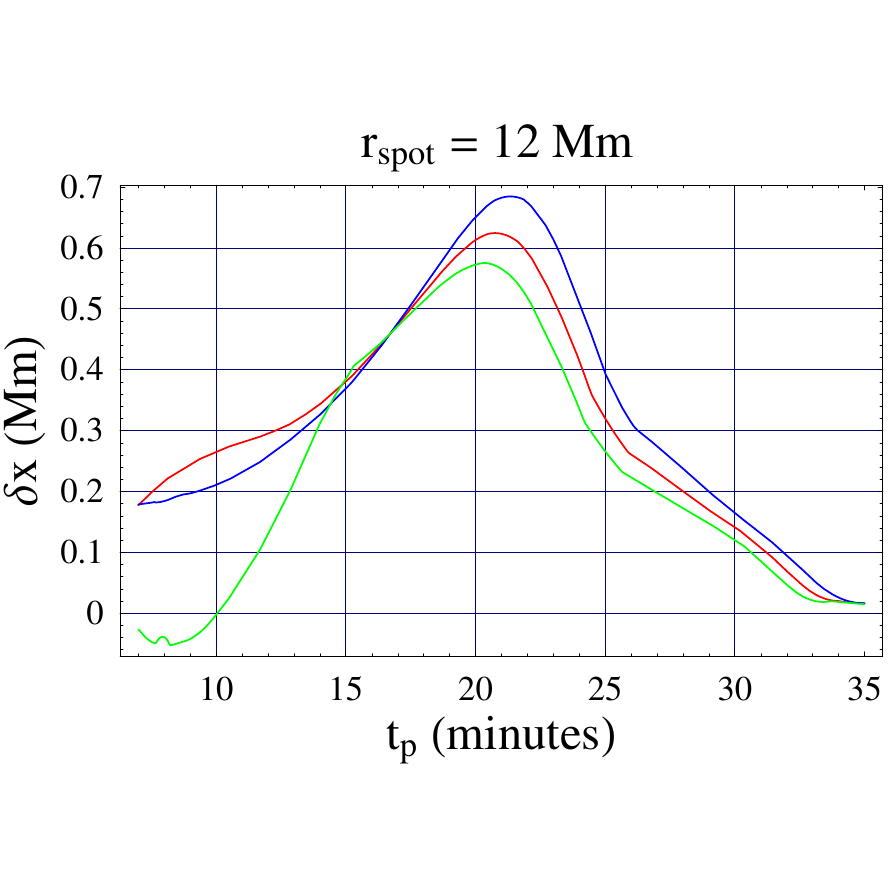}&
\hspace*{-3mm}
\includegraphics[trim= 0mm 12mm 0mm 12mm, clip, width=13.5pc]{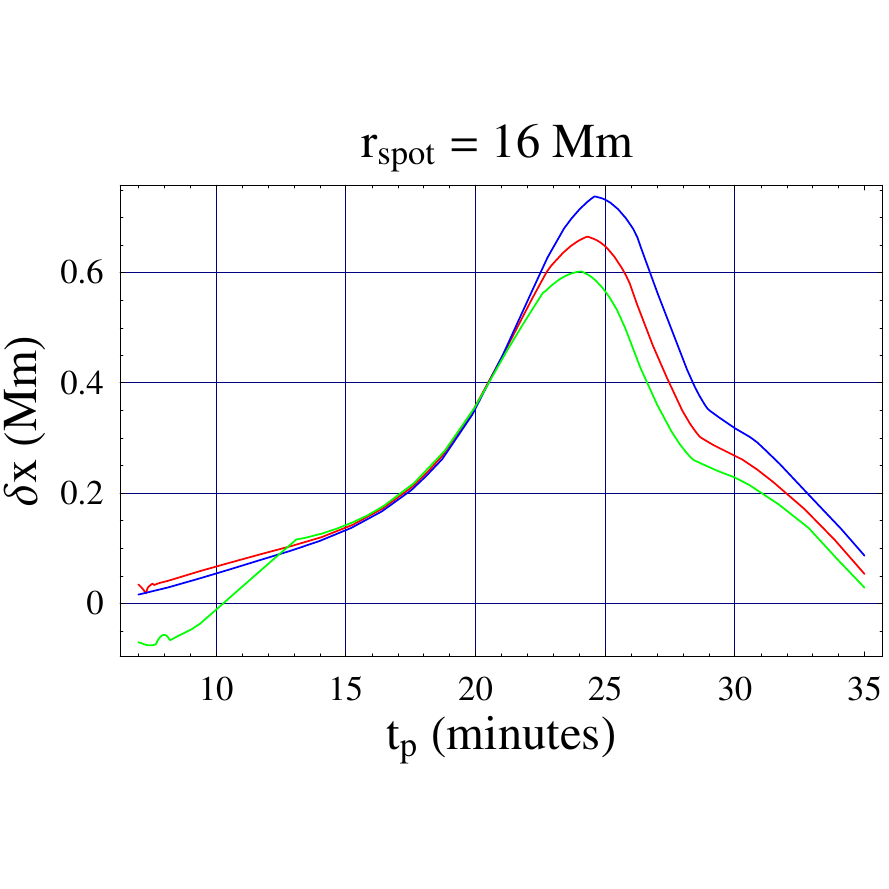}
\end{tabular}
\end{center}
\caption{Skip distance perturbations ($\updelta_x$) as a function of phase travel time ($t_{p}$) for $r_{\mathrm{spot}}=4,8,12$, and $16$~Mm on the sunspot, calculated for three frequencies$\omega=3.5$ (green), $\omega=4$ (red), and $\omega=5$~mHz (blue).}
\label{fig:deltax}
\end{figure}

Furthermore, in Figure~\ref{fig:deltax} we can see that there are also significant skip-distance perturbations ($\updelta x$) associated with rays that are propagated through the sunspot atmosphere. These calculations are for similar positions and frequencies as in Figure~\ref{fig:deltat}. The exclusively positive values of $\updelta x$ that we can see along the sunspot radius indicates that at the same time that these rays are being sped up, they are also undertaking a longer journey than their Model S counterparts in the process, and as with $\updelta\tau_\mathrm{p}$, the magnitude of the calculated $\updelta x$ appears to be closely related to surface magnetic field strength. For both $\updelta\tau_\mathrm{p}$ and $\updelta x$ we also observe a particular pattern of perturbation associated with each position along the sunspot. Whereas the perturbations appear to mainly decrease when we are close to spot centre (\textit{e.g.} $r_{\mathrm{spot}}=4, 8$~Mm), they appear to increase when further away (\textit{e.g.} $r_{\mathrm{spot}}=12, 16$~Mm) from spot centre. This is clearly a bi-product of both varying field strength and inclination angle of field lines (see Figure~\ref{fig:brbz}) as we move across the sunspot. Field strength tends to decrease, while field lines become more significantly inclined as we move away from centre of the sunspot. 

Also clearly obvious from both Figures~\ref{fig:deltat} and \ref{fig:deltax} is the presence of a significant frequency dependence of both $\updelta\tau_\mathrm{p}$ and $\updelta x$  measurements in the sunspot, with the magnitudes of the perturbations increasing as the frequency is increased from $3.5$ to $5$~mHz. This is particularly evident for rays with short skip distances (\textit{i.e.} surface skimmers with very shallow lower turning points). Frequency dependence of travel-time perturbations in active regions has also been observed by both helioseismic holography \cite{bb2006} and time-distance helioseismology \cite{sebraj}. We shall discuss the importance of these observations in greater detail in the upcoming sections. \inlinecite{cally06} also observed a similar behaviour when modelling rays in inclined fields and described several related but distinct effects that strong magnetic fields appear to have on seismic waves, with an important ``dual effect'' that the magnetic field has on individual ray paths (that is, increasing their skip distances while at the same time, speeding them up considerably) being one of these effects. 
\begin{figure}[h]
\begin{center}
\begin{tabular}{c}
\includegraphics[trim= 0mm 58mm 0mm 58mm, clip, width=23.45pc]{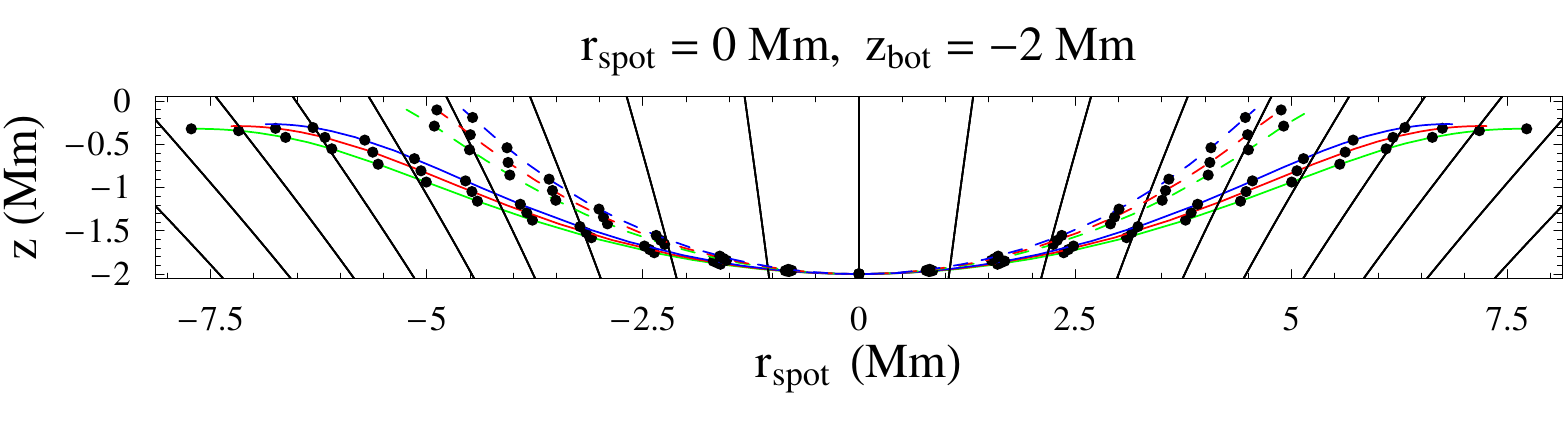}\\
\includegraphics[trim= 0mm 58mm 0mm 58mm, clip, width=23.45pc]{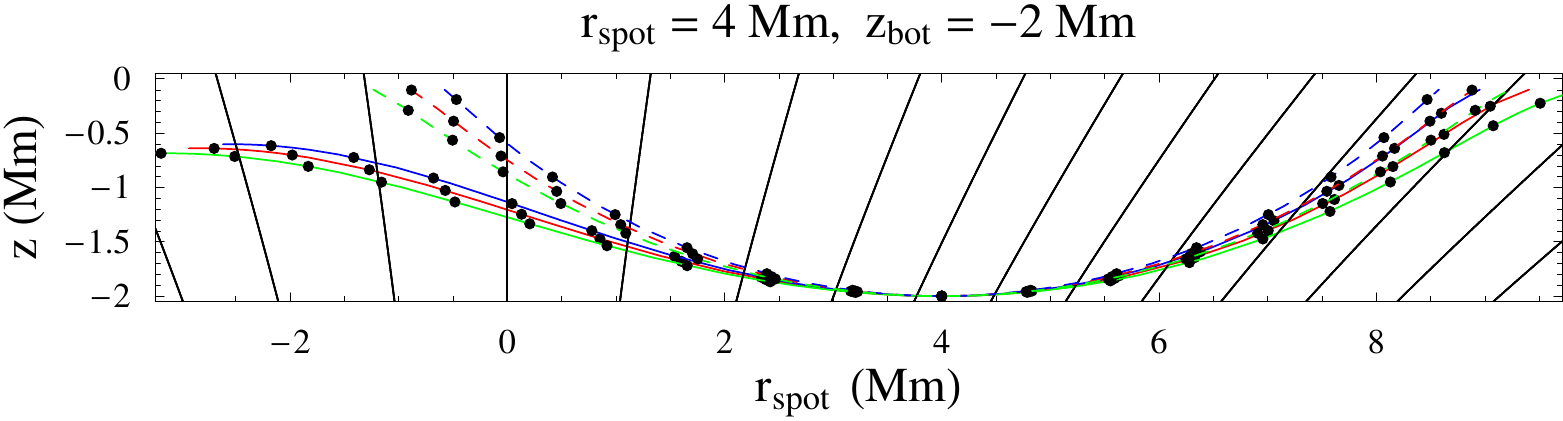}\\
\includegraphics[trim= 0mm 56mm 0mm 56mm, clip, width=23.45pc]{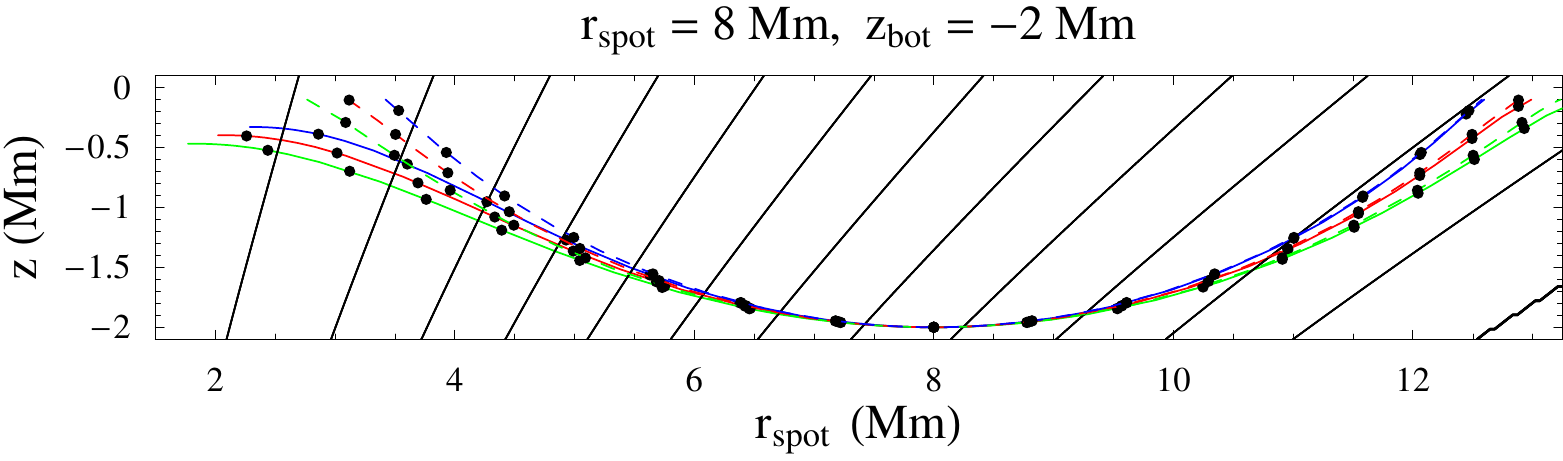}\\
\includegraphics[trim= 0mm 56mm 0mm 56mm, clip, width=23.45pc]{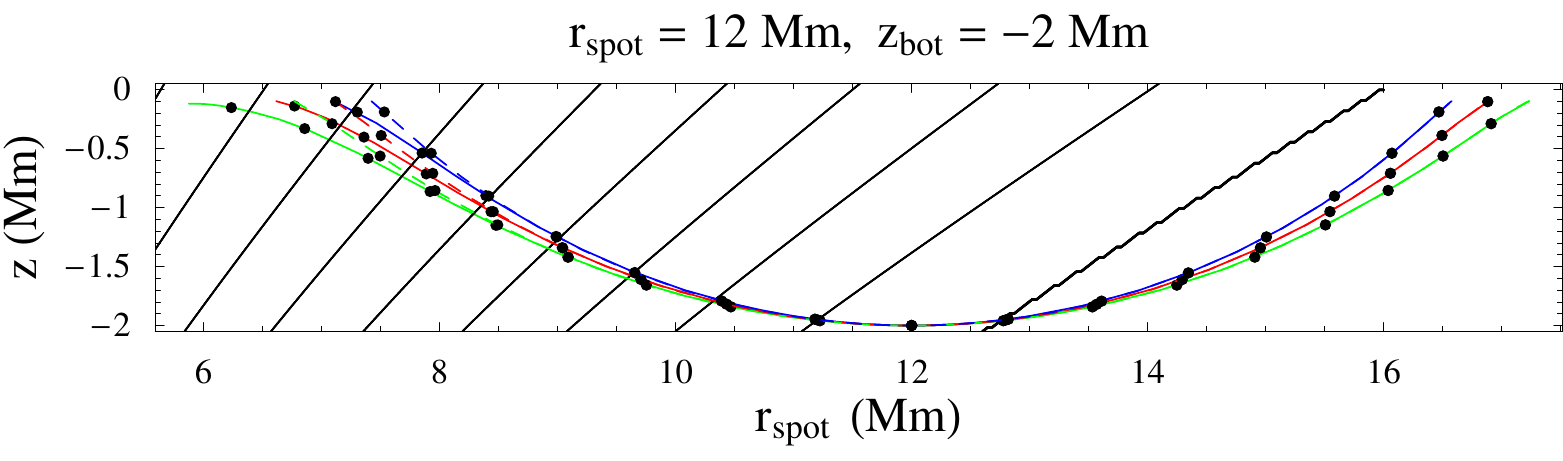}\\
\includegraphics[trim= 0mm 56mm 0mm 56mm, clip, width=23.45pc]{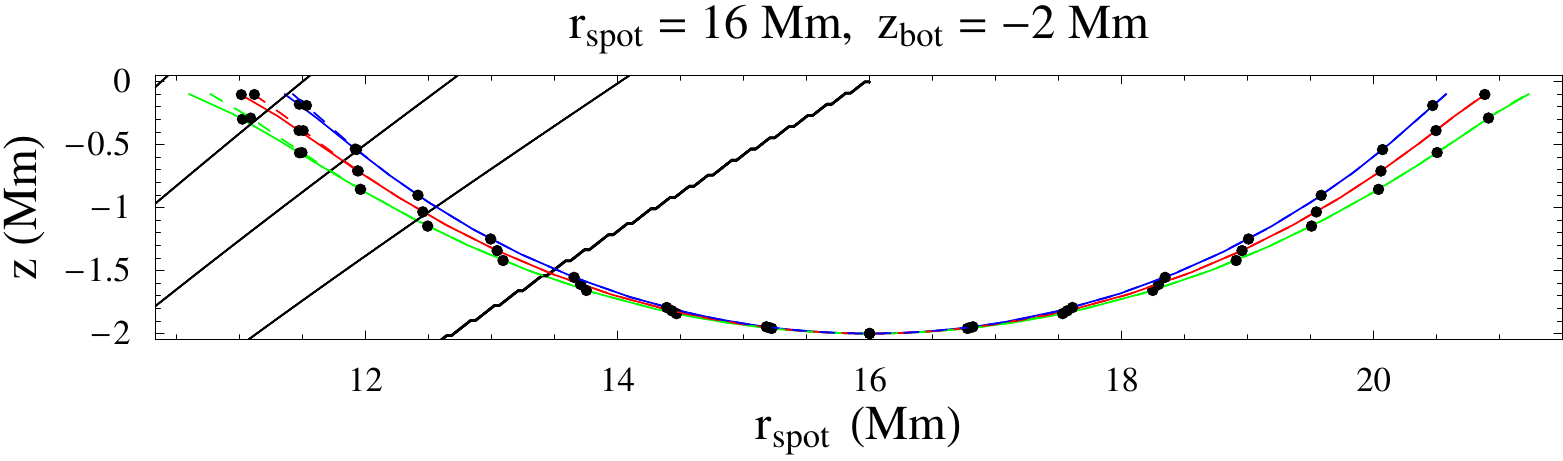}
\end{tabular}
\end{center}
\caption{Individual rays propagated through the simulated sunspot (solid rays) and Model S (dashed rays) atmospheres, calculated for three frequencies: $\omega=3.5$ (green), $\omega=4$ (red), and $\omega=5$~mHz (blue). The top of each frame indicates the initial depth ($z_\mathrm{bot}$, Mm) and radial grid position of the lower turning point of the ray ($r_\mathrm{spot}$, Mm). }  
\label{fig:rays}
\end{figure}

A comparison between rays propagated inside the sunspot model with rays propagated in the quiet-Sun clearly reveals these effects to the naked eye. All rays shown in Figure~\ref{fig:rays} are initialized at a depth of $z_\mathrm{bot}=-2$~Mm, with the rays inside the sunspot model (solid rays, colours identify frequencies) also being initialized at varying positions along the sunspot ($r_\mathrm{spot}=0, 4, 8, 12$, and $16$~Mm). While the rays propagated inside the Model S atmosphere (dashed rays) are symmetrical about their turning points (as expected), strong asymmetries (at both turning points) are associated with the same rays when initiated inside the sunspot. We can clearly see that the rays inside the sunspot (at all three frequencies) appear to have undergone a longer skip distance, in a slightly shorter amount of time (dots along ray paths indicate one-minute $t_\mathrm{g}$ intervals), confirming the perturbation profiles of Figures~\ref{fig:deltat} and \ref{fig:deltax}. Of course Figure~\ref{fig:rays} shows a very small sample of rays initialized at a given depth, but even so, they are quite clearly indicative of the large-scale effects of the magnetic field on ray propagation -- effects which are more pronounced as we approach the spot centre and in regions of significantly inclined magnetic fields.

\subsection{Binned Travel-Time Perturbation Profiles}
\begin{figure}[ht]
\begin{center}
\begin{tabular}{ccc}
\hspace*{-5mm}
\includegraphics[trim= 9mm 0mm 9mm 0mm, clip,width=9.5pc]{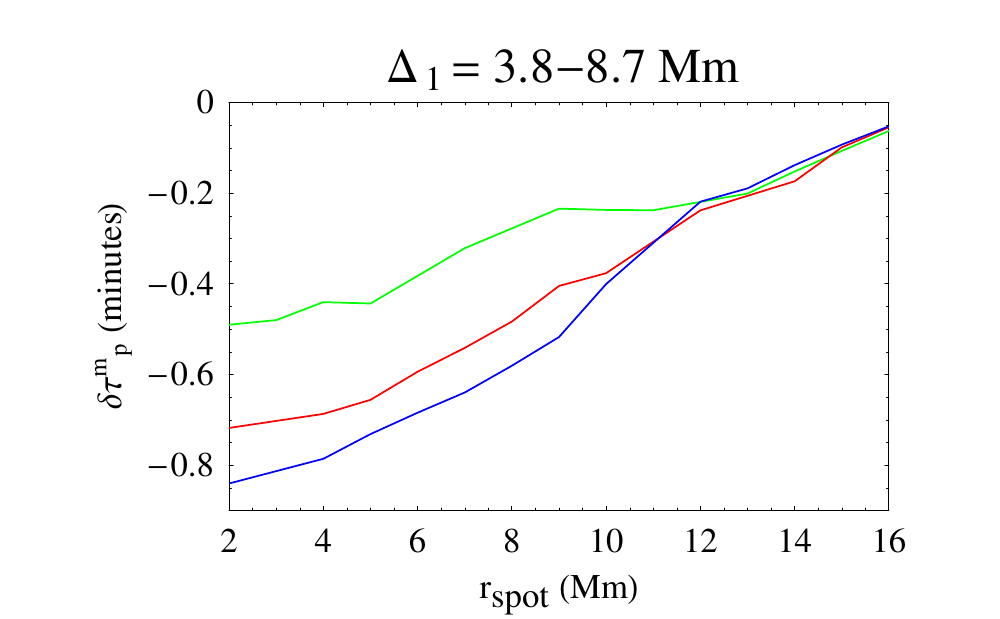}&
\hspace*{-5mm}
\includegraphics[trim= 9mm 0mm 9mm 0mm, clip,width=9.5pc]{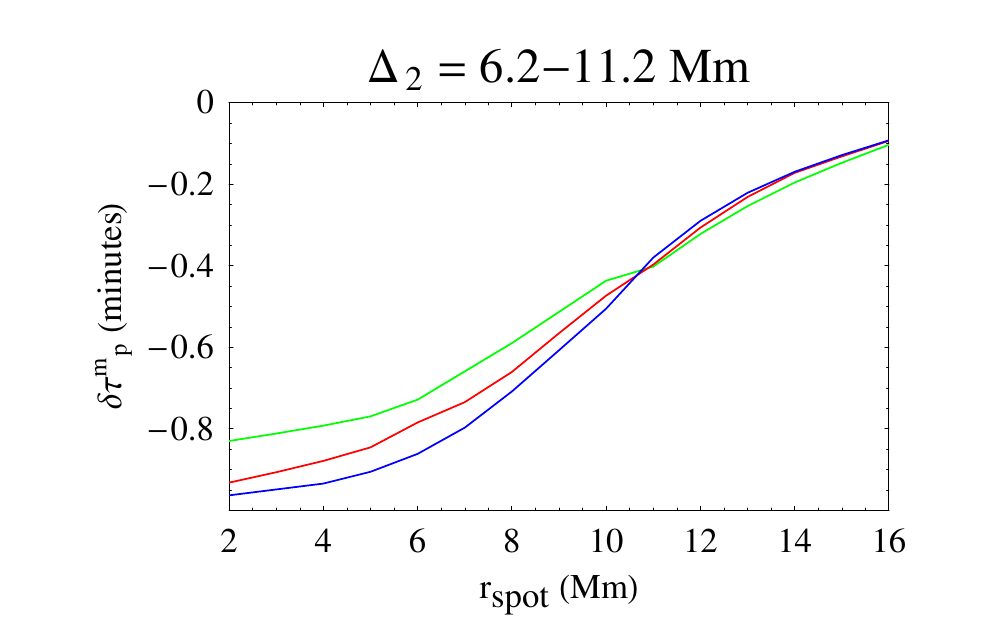}&
\hspace*{-5mm}
\includegraphics[trim= 9mm 0mm 9mm 0mm, clip,width=9.5pc]{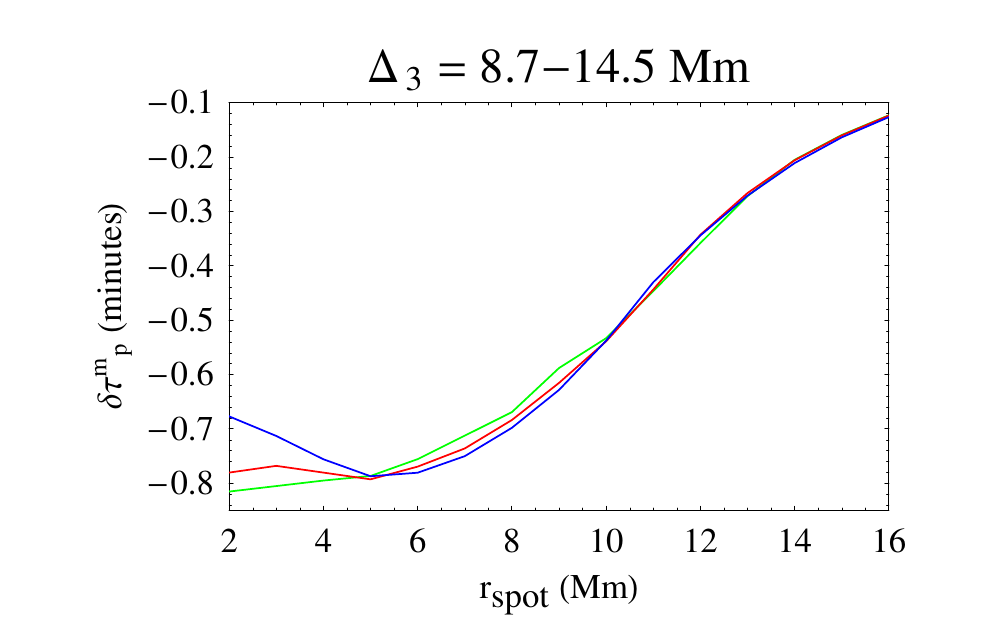}\\
\hspace*{-5mm}
\includegraphics[trim= 9mm 0mm 9mm 0mm, clip,width=9.5pc]{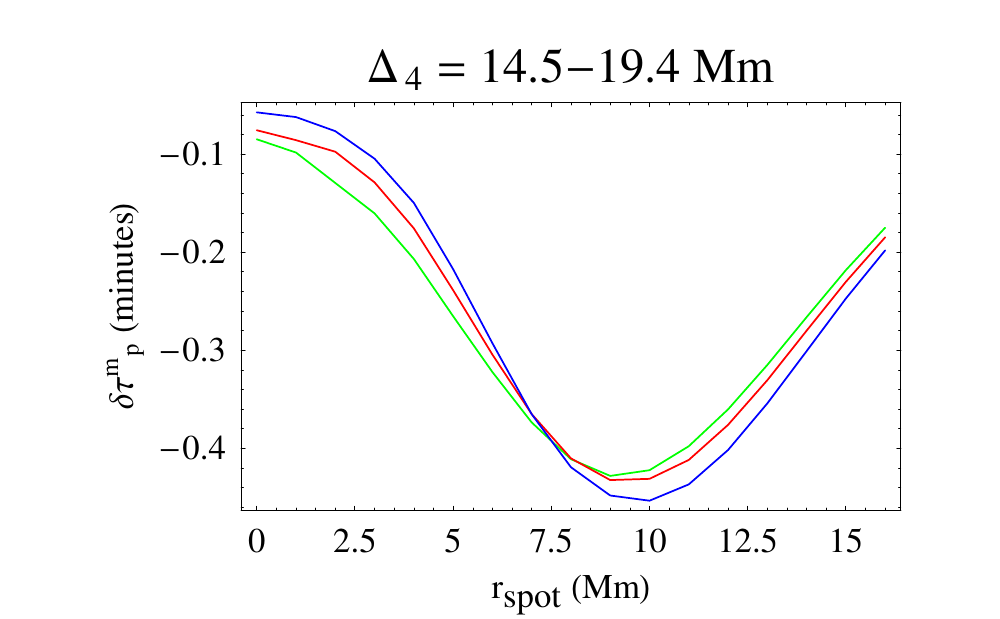}&
\hspace*{-5mm}
\includegraphics[trim= 7mm 0mm 8mm 0mm, clip,width=9.5pc]{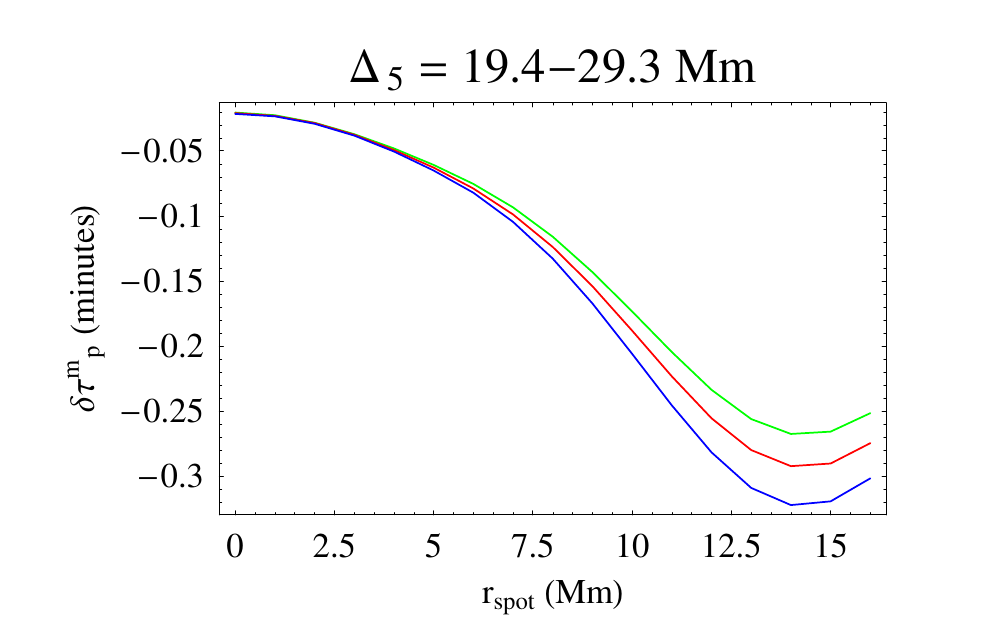}&
\hspace*{-5mm}
\includegraphics[trim= 7mm 0mm 8mm 0mm, clip,width=9.5pc]{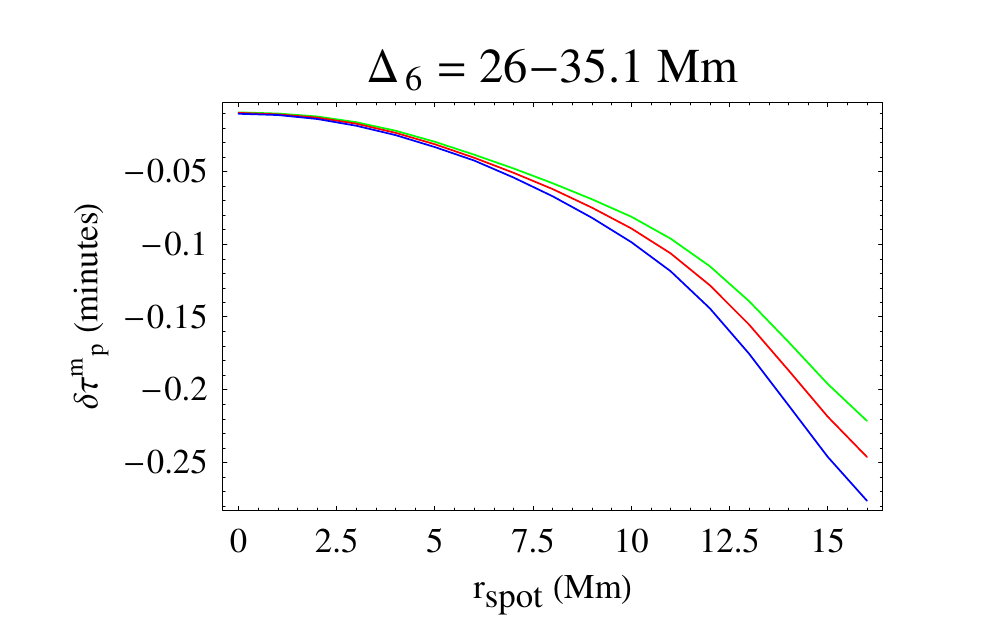}\\
\hspace*{-5mm}
\includegraphics[trim= 6mm 0mm 8mm 0mm, clip,width=9.5pc]{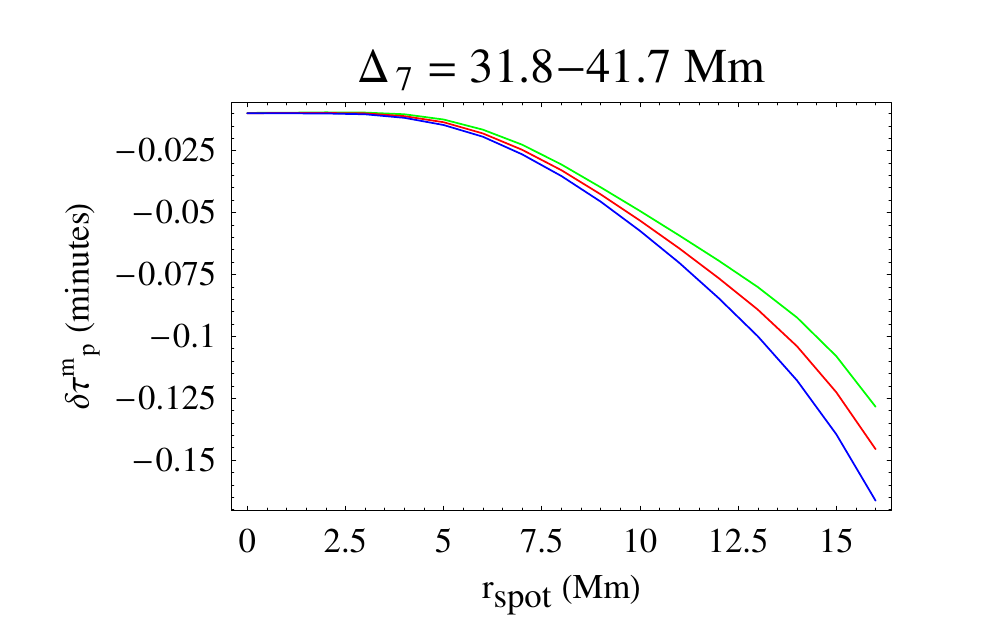}&
\hspace*{-5mm}
\includegraphics[trim= 7mm 0mm 8mm 0mm, clip,width=9.5pc]{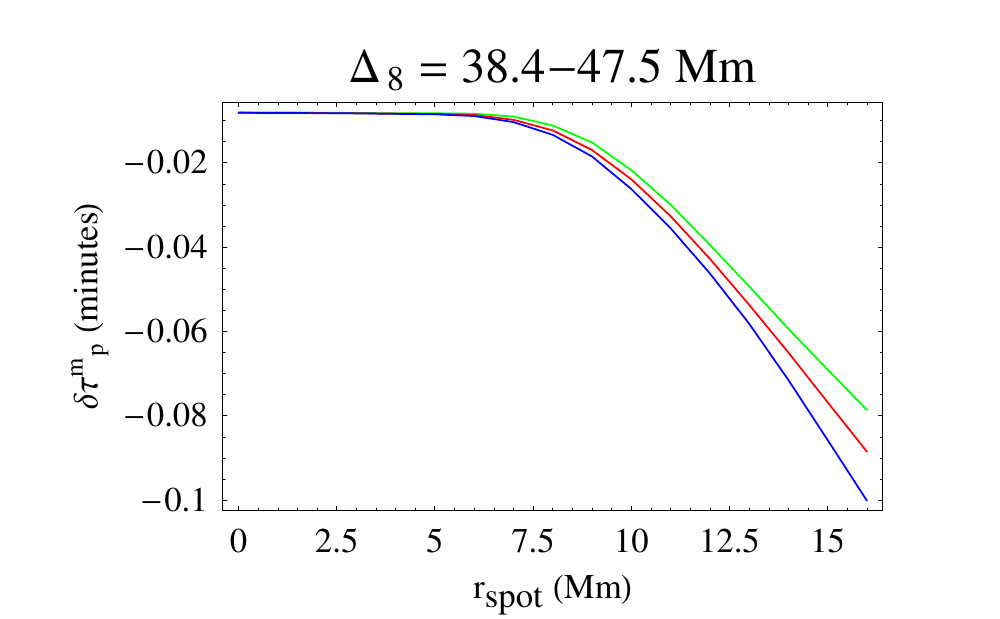}&
\hspace*{-5mm}
\includegraphics[trim= 7mm 0mm 8mm 0mm, clip,width=9.5pc]{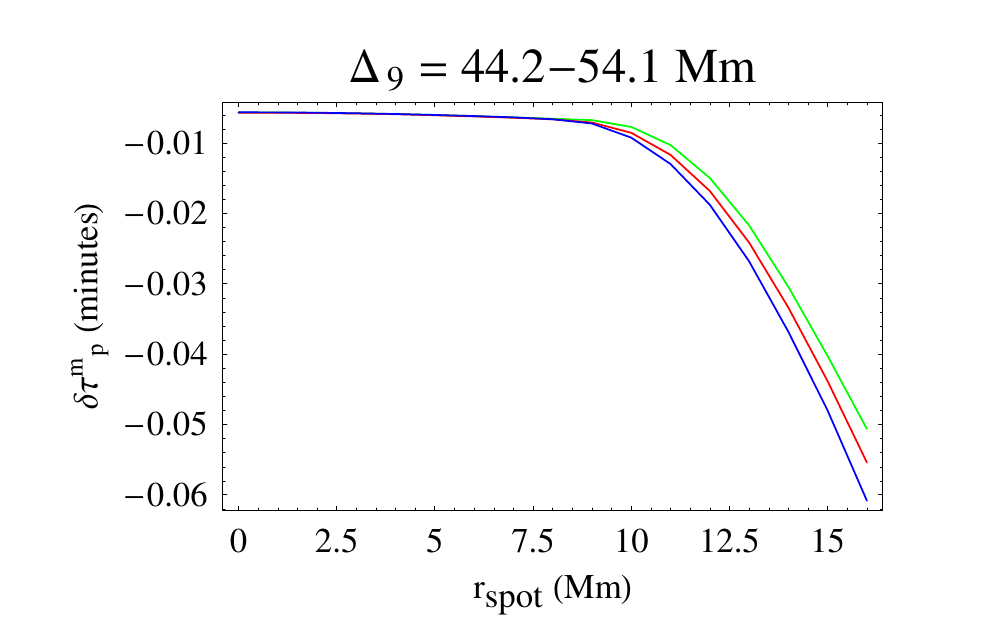}\\
\hspace*{-5mm}
\includegraphics[trim= 6mm 0mm 8mm 0mm, clip,width=9.5pc]{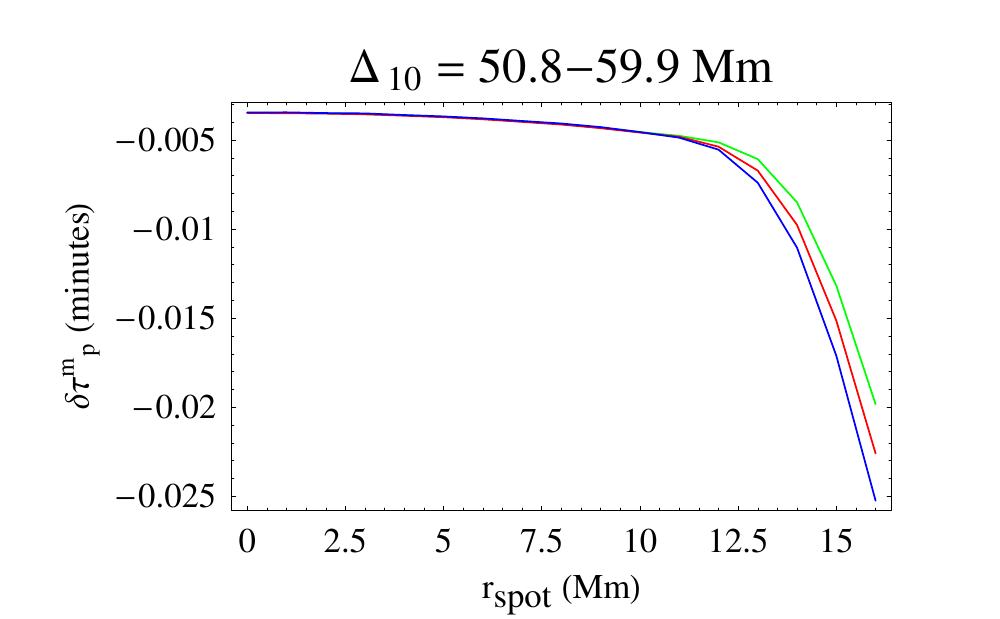}&
\hspace*{-5mm}
\includegraphics[trim= 6mm 0mm 8mm 0mm, clip,width=9.5pc]{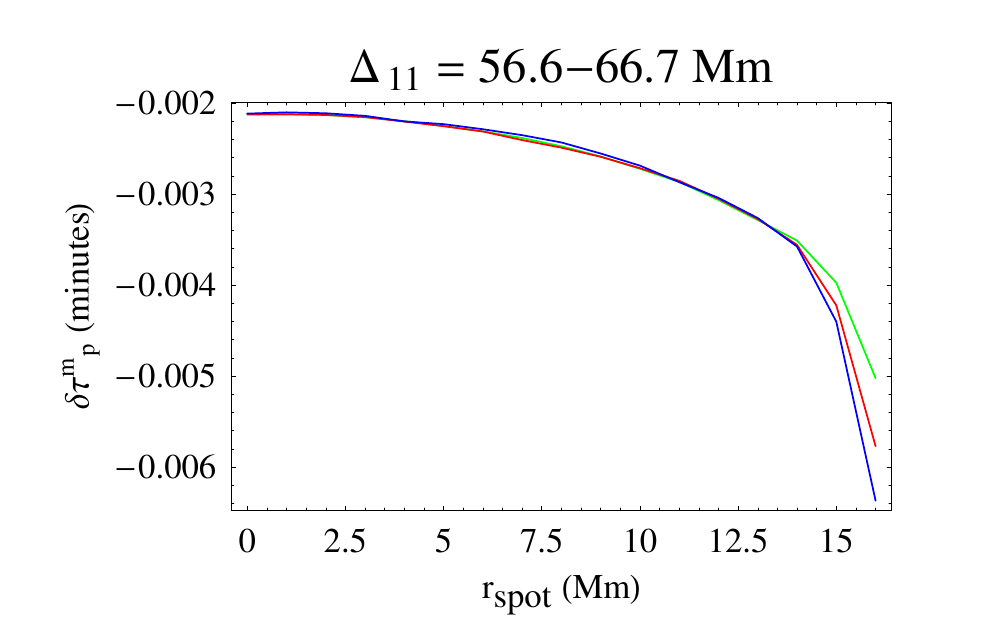}
\end{tabular}
\end{center}
\caption{Binned (mean) travel-time perturbation ($\updelta\tau^\mathrm{m}_\mathrm{p}$, minutes) profiles  as a function of position ($r_\mathrm{spot}$, Mm) on the sunspot, calculated for three frequencies: $\omega=3.5$ (green), $\omega=4$ (red), and $\omega=5$~mHz (blue). Annuli number and sizes are indicated on the top of the frame of each bin.}
\label{fig:bindt}
\end{figure}

The mean ray travel-time perturbations ($\updelta\tau^\mathrm{m}_\mathrm{p}$) for each frequency and grid position were calculated and binned into 11 annuli ($\Delta_1 - \Delta_{11}$) of various sizes (outlined in Table~\ref{tab:bins}). The $\updelta\tau^\mathrm{m}_\mathrm{p}$ profiles of the bins are shown in Figure~\ref{fig:bindt}. Once again, we can see the clear frequency dependence of travel-time perturbations evident in all bins, with perturbations increasing with increasing frequency as before. Also, all $\updelta\tau^\mathrm{m}_\mathrm{p}$ bins contain negative perturbations as we saw before in Figure~\ref{fig:deltat}. We also observe that the magnitude of $\updelta\tau^\mathrm{m}_\mathrm{p}$ decreases as we move away from the centre of the sunspot (\textit{i.e.} decreasing field strength) for the smaller bins (\textit{e.g.} $\Delta_1 - \Delta_{3}$). 

These smaller bins are representative of shallow rays that spend a considerable proportion of their journey inside the magnetic field, consistent with the larger magnitude of the perturbations seen in these bins. Larger bins (\textit{e.g.} $\Delta_4 - \Delta_{11}$) sample rays with much deeper lower turning points, hence a considerable amount of the journey undertaken by these rays would be spent in the quiet-Sun Model S atmosphere. Therefore the magnitude of the perturbations tends to be smaller than that for the smaller bins. However, they are found to increase in magnitude as we move away from the centre of the sunspot as rays sample larger areas of the magnetic field throughout their journey across the sunspot radius.

It should be noted that for the smaller bins (particularly for $\Delta_1 - \Delta_3$), it becomes quite difficult to obtain a sufficient sampling of rays to average near the centre of the flux tube, even with a very fine grid spacing of $\Delta z=-0.025$~Mm in the very sensitive top $1.5$~Mm of the computational grid. As such, we get a certain level of rigidity in the $\updelta\tau^\mathrm{m}_\mathrm{p}$ profiles of these bins. No such restriction is encountered when using a pure Model S atmosphere, which tends to suggest that strong near-surface magnetic fields are severely restricting the propagation of helioseismic rays with short skip distances (or very shallow lower turning points). 
 
Although our sunspot model has many of the qualitative features we might expect in a real spot, it is nonetheless rather \emph{ad hoc}, and consequently our time-distance results do not warrant detailed comparison with solar observations. Nevertheless, it is of interest to qualitatively compare the $\updelta\tau^\mathrm{m}_\mathrm{p}$ results obtained from our simulations to those reported for AR 8243 (18 June 1998) by \inlinecite{cbk06}. S. Couvidat kindly provided us with the actual set of travel time maps used in their analysis. 

To compare the $\updelta\tau^\mathrm{m}_\mathrm{p}$ profiles as closely as possible, we first compute the azimuthal average of the four $\updelta\tau^\mathrm{m}_\mathrm{p}$ maps presented in Figure 3 of \inlinecite{cbk06} (corresponding to $\Delta_1$, $\Delta_3$, $\Delta_6$ and $\Delta_9$, noting that the travel-times were obtained without a frequency bandpass filter), to obtain $\updelta\tau^\mathrm{m}_\mathrm{p}$ profiles of AR 8243, akin to our artificial $\updelta\tau^\mathrm{m}_\mathrm{p}$ profiles contained in Figure~\ref{fig:bindt}. We observe peak (positive) travel-time perturbations of $\approx0.29$ and $\approx0.16$ minutes respectively for $\Delta_1$ and $\Delta_3$ in the sunspot umbra, while the sign of $\updelta\tau^\mathrm{m}_\mathrm{p}$ in the sunspot changes for the larger bins, $\Delta_6$ and $\Delta_9$, with $\updelta\tau^\mathrm{m}_\mathrm{p}$ ranging from $\approx -0.38$ to $\approx -0.31$ minutes respectively. The perturbations for all four bins also appear to decrease in the penumbra relative to the umbra. In comparison, if we assume a central frequency of 3.5~mHz, the artificial $\updelta\tau^\mathrm{m}_\mathrm{p}$ profiles for the bins produced by our simulations (Figure~\ref{fig:bindt}, 3.5~mHz profiles indicated by solid green lines) show opposite-in-sign and larger-in-magnitude $\updelta\tau^\mathrm{m}_\mathrm{p}$ for both $\Delta_1$ ($\approx-0.7$ minutes) and $\Delta_3$ ($\approx-0.82$ minutes), while similar-in-sign yet smaller-in-magnitude $\updelta\tau^\mathrm{m}_\mathrm{p}$ profiles were observed for $\Delta_6$ ($\approx-0.22$ minutes) and $\Delta_9$ ($\approx-0.05$ minutes). When we consider higher frequencies, the magnitude of the artificial $\updelta\tau^\mathrm{m}_\mathrm{p}$ increases with frequency for all four bins, with all perturbations being negative in sign. However, the general pattern of the artificial $\updelta\tau^\mathrm{m}_\mathrm{p}$ profiles for all frequencies appears to be similar to the observations of \inlinecite{cbk06}, with perturbations decreasing with increasing radius from the centre of the sunspot. 

While the differences in the magnitudes of $\updelta\tau^\mathrm{m}_\mathrm{p}$ between our simulations and those of \inlinecite{cbk06} (at a given fixed central frequency) can be explained, to some extent, by magnetic and thermal differences between our model and their sunspot, the frequency dependence of $\updelta\tau^\mathrm{m}_\mathrm{p}$ and the sign change of the smaller bins in particular (\textit{i.e.} positive $\updelta\tau^\mathrm{m}_\mathrm{p}$ resulting from actual time-distance observations, negative $\updelta\tau^\mathrm{m}_\mathrm{p}$ from the simulations) can not be dismissed as easily. Traditionally, positive $\updelta\tau^\mathrm{m}_\mathrm{p}$ obtained for short skip distances in sunspots have been interpreted as representing a region of slower wave-speed propagation in the shallow sub-surface layers of the sunspot. However, as we briefly noted in the previous section, \inlinecite{bb2006} (using helioseismic holography) found that, at a given fixed phase speed, travel-time perturbations within active regions exhibit a strong frequency dependence. \inlinecite{sebraj} confirmed these results using time-distance helioseismology, applying additional frequency bandpass filters (centred at 3, 4 and 4.5~mHz) to the standard phase-speed filters used in \inlinecite{cbk06} in order to determine the cause of the dark rings of negative $\updelta\tau^\mathrm{m}_\mathrm{p}$ they detected in the travel-time maps (mainly associated with the $\Delta_2$ and $\Delta_3$ skip-distance bins) of a majority of the sunspots they studied. These rings, which are sensitive to the frequency filtering applied, are found to produce significant ring-like structures in the inversion results, mimicking regions of increased sound speed. The authors conclude that the rings are most likely to be artifacts caused by surface effects, probably of magnetic origin. 

In addition to these results, the very recent work undertaken by \inlinecite{bb08} (using ridge filters, in addition to the standard phase-speed filters) provide strong evidence that the positive perturbations observed arise from the $p_1$ ridge or beneath it. These positive travel-time shifts were not seen in the higher order $p$-mode data. These results, when considered in conjunction with our artificial $\updelta\tau^\mathrm{m}_\mathrm{p}$ profiles (and the results contained in in the next section), provide further concrete evidence that positive travel-time perturbations obtained for short skip distances are likely to be artifacts or bi-products of the data reduction or analysis method used, rather than some actual physical sub-surface anomaly below the sunspot.  

\subsection{Isolating the Thermal Component of Travel Time Perturbations}
One of the keys to understanding the role played by near-surface magnetic fields in local helioseismology is to be able to isolate it from effects thought to be produced by thermal or flow perturbations. The simplest way to isolate such effects is to essentially ``switch off'' the magnetic field when calculating the ray paths in the simulations -- that is, set $a=0$ in the simulated sunspot atmosphere, but maintain the modified sound-speed profile obtained (seen in Figure~\ref{fig:thermals}). 
\begin{figure}[ht]
\begin{center}
\begin{tabular}{ccc}
\hspace*{-5mm}
\includegraphics[trim= 9mm 0mm 9mm 0mm, clip,width=9.5pc]{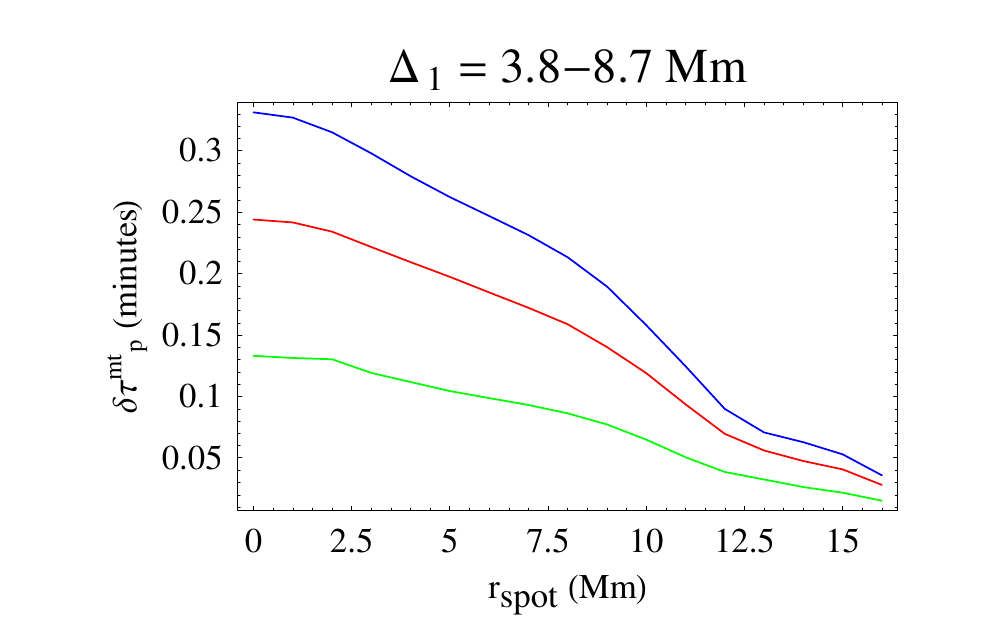}&
\hspace*{-5mm}
\includegraphics[trim= 9mm 0mm 9mm 0mm, clip,width=9.5pc]{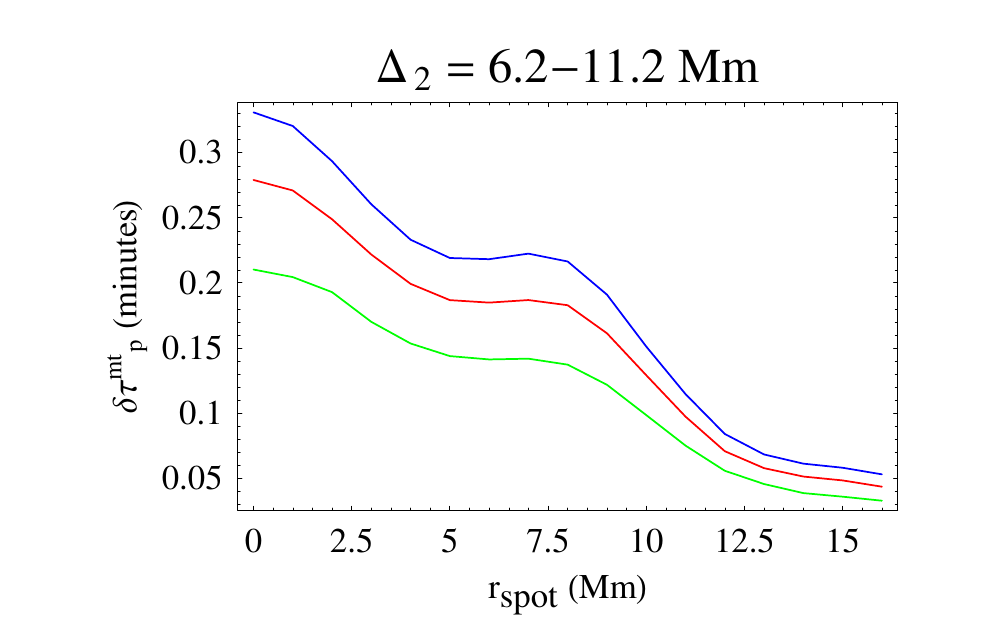}&
\hspace*{-5mm}
\includegraphics[trim= 9mm 0mm 9mm 0mm, clip,width=9.5pc]{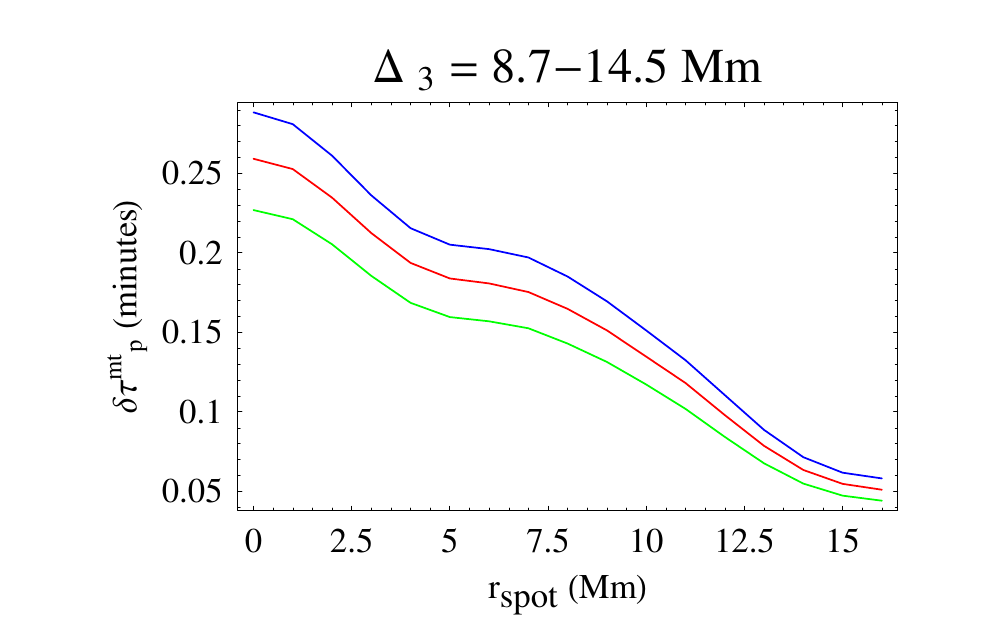}\\
\hspace*{-5mm}
\includegraphics[trim= 9mm 0mm 9mm 0mm, clip,width=9.5pc]{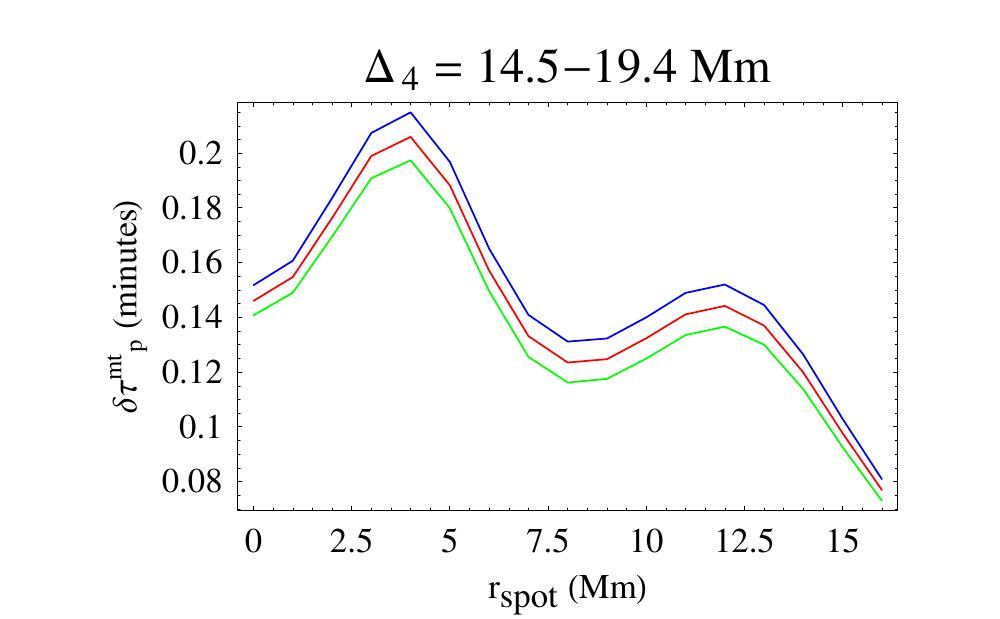}&
\hspace*{-5mm}
\includegraphics[trim= 7mm 0mm 8mm 0mm, clip,width=9.5pc]{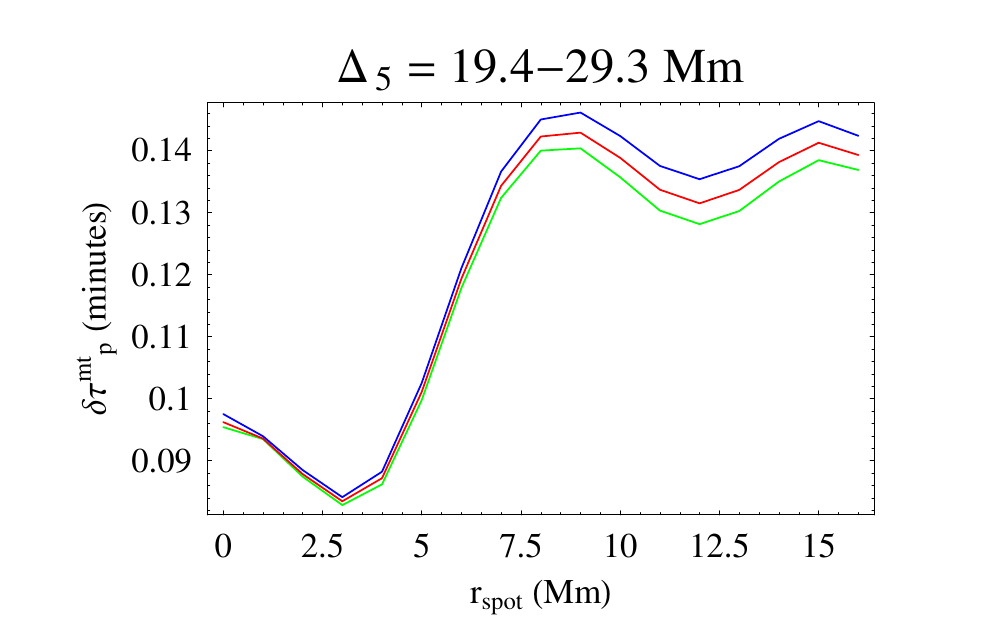}&
\hspace*{-5mm}
\includegraphics[trim= 7mm 0mm 8mm 0mm, clip,width=9.5pc]{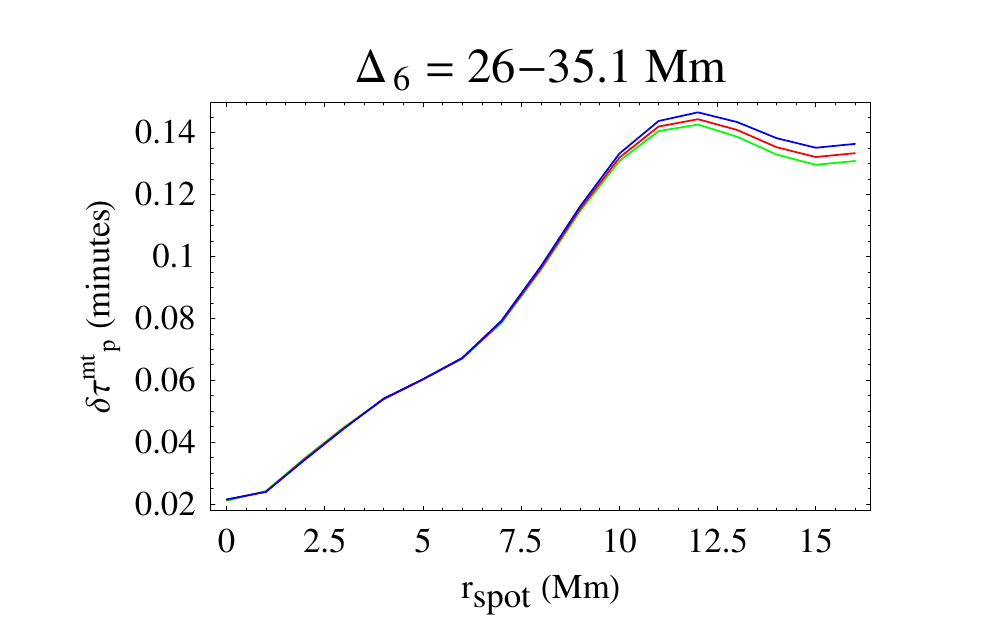}\\
\hspace*{-5mm}
\includegraphics[trim= 6mm 0mm 8mm 0mm, clip,width=9.5pc]{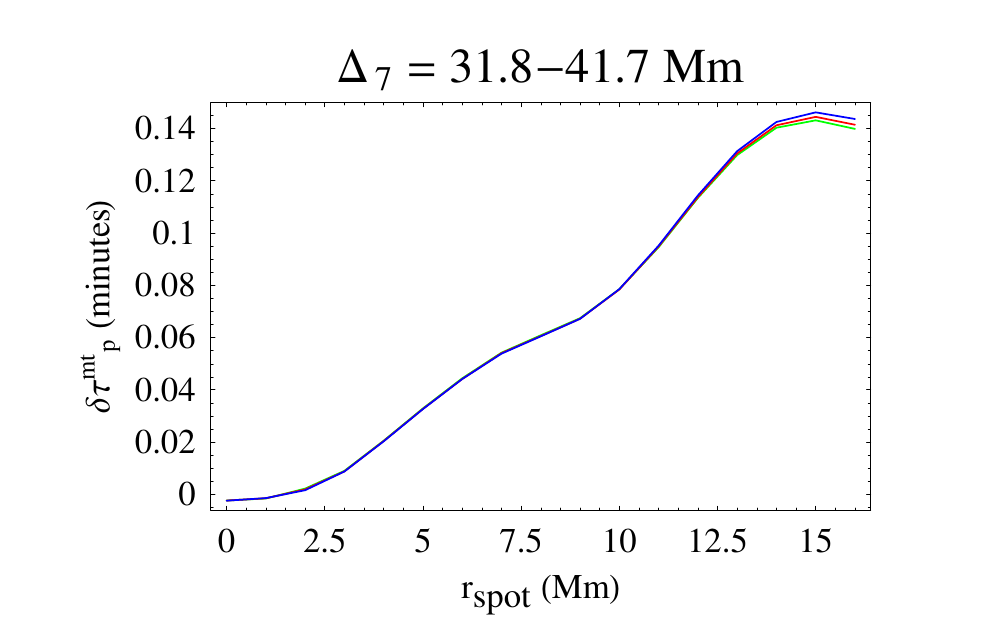}&
\hspace*{-5mm}
\includegraphics[trim= 7mm 0mm 8mm 0mm, clip,width=9.5pc]{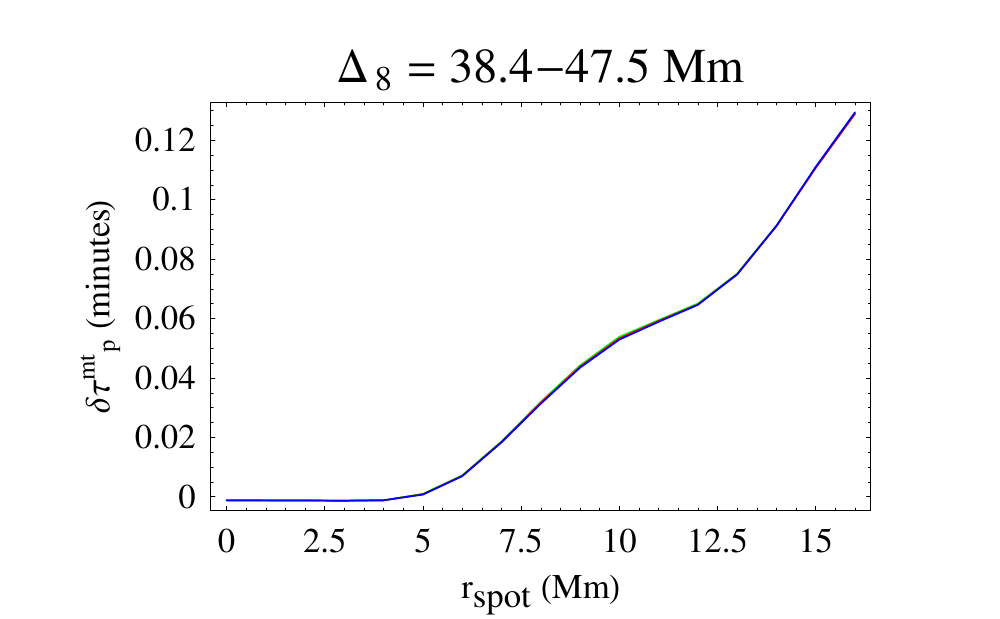}&
\hspace*{-5mm}
\includegraphics[trim= 7mm 0mm 8mm 0mm, clip,width=9.5pc]{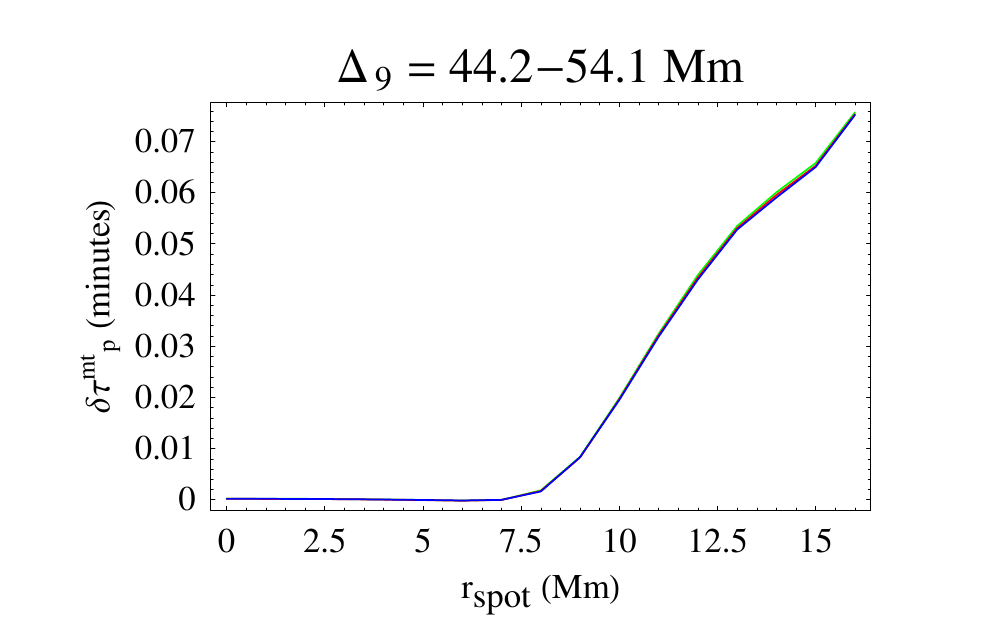}\\
\hspace*{-5mm}
\includegraphics[trim= 6mm 0mm 8mm 0mm, clip,width=9.5pc]{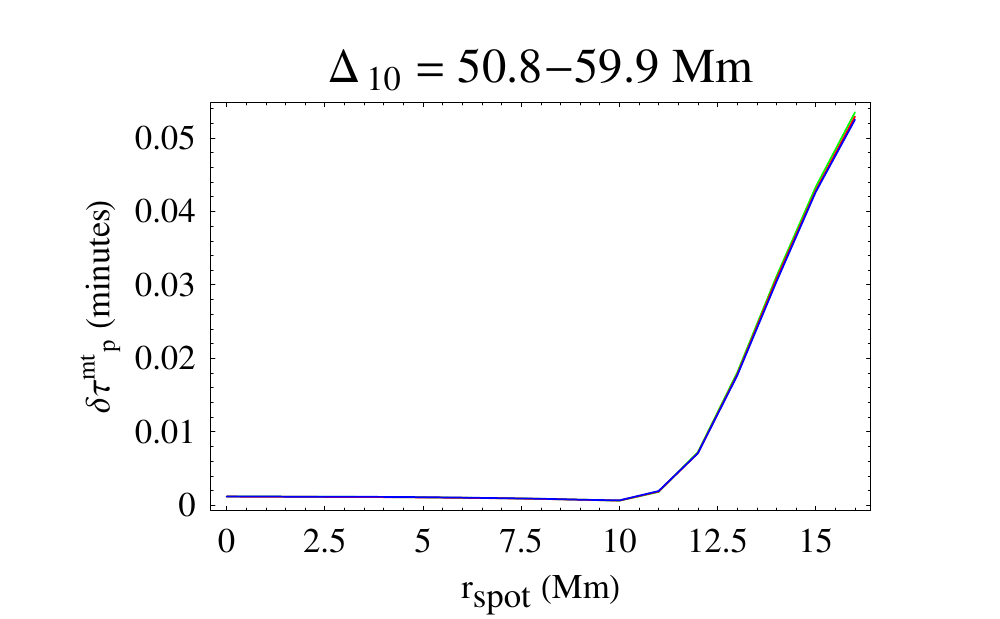}&
\hspace*{-5mm}
\includegraphics[trim= 6mm 0mm 8mm 0mm, clip,width=9.5pc]{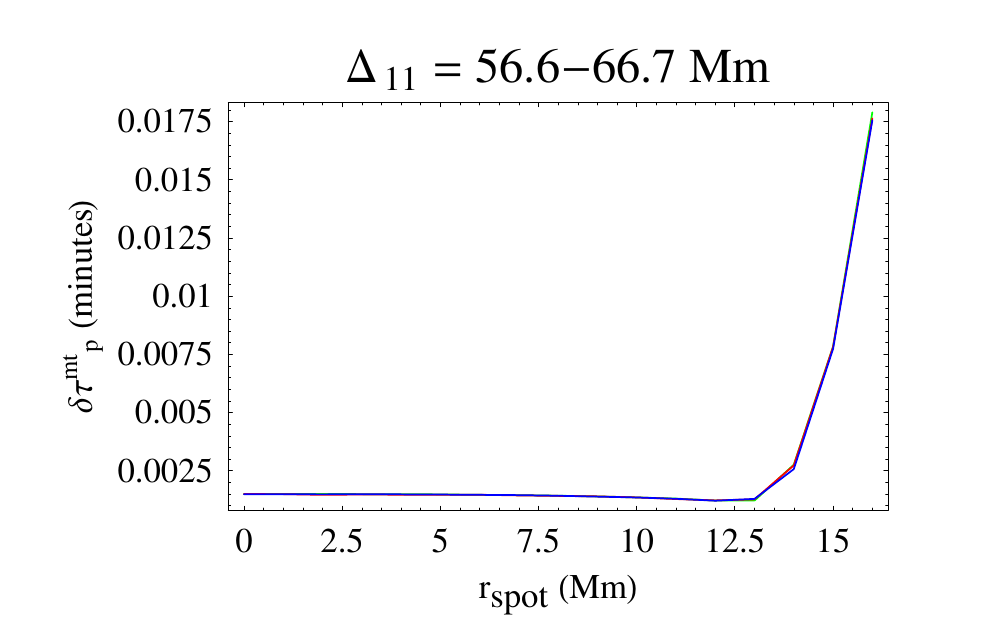}
\end{tabular}
\end{center}
\caption{Binned (mean) thermal travel-time perturbation ($\updelta\tau^\mathrm{mt}_\mathrm{p}$, minutes) profiles as a function of position ($r_\mathrm{spot}$, Mm) on the sunspot, calculated for three frequencies: $\omega=3.5$ (green), $\omega=4$ (red), and $\omega=5$~mHz (blue). Annuli number and sizes are indicated on the top of the frame of each bin.}
\label{fig:bindt0}
\end{figure}

The external atmosphere, ray-path simulations and computational grid remain identical to those described previously. The only difference is the resulting thermal travel-time perturbations ($\updelta\tau^\mathrm{mt}_\mathrm{p}$) which would then be purely a result of what can be referred to as ``thermal variations'' along the ray path. One can then compare the resulting perturbation profiles to those obtained when the magnetic field is included in the simulations (\textit{i.e.} Figure~\ref{fig:bindt}) to better understand the role of the thermal contributions to the observed $\updelta\tau^\mathrm{m}_\mathrm{p}$ profiles. Figure~\ref{fig:bindt0} shows the resulting bins of the thermal component of $\updelta\tau^\mathrm{mt}_\mathrm{p}$.   

In general, the resulting $\updelta\tau^\mathrm{mt}_\mathrm{p}$ profiles are relatively smooth and all bins clearly show exclusively positive travel-time perturbations (compared to exclusively negative travel-time perturbations observed in Figure~\ref{fig:bindt}), this implies that rays are travelling considerably slower than in the Model S atmosphere -- a clear contrast with simulations where the magnetic field is present. The magnitude of $\updelta\tau^\mathrm{mt}_\mathrm{p}$ is also decreasing with increasing radius for the smaller bins ($\Delta_1 - \Delta_4$) and \textit{vice versa} for the larger bins ($\Delta_5 - \Delta_{11}$), a similar behaviour to what is observed in Figure~\ref{fig:bindt}. However, when considering the magnitude of the perturbations between Figures~\ref{fig:bindt} and \ref{fig:bindt0}, it is clear that thermal perturbations appear to be much smaller for a majority of the bins -- in fact up to 400\% smaller for some frequencies when comparing the perturbations in the near-surface regions $\Delta_1 - \Delta_{3}$. The magnitude of the perturbations become much more comparable when looking at the larger bins ($\Delta_7$ onwards), and from $\Delta_8$ onwards $\updelta\tau^\mathrm{mt}_\mathrm{p}$ becomes ever slightly larger than the ones we see in Figure~\ref{fig:bindt} for the same bins. Frequency dependence of $\updelta\tau^\mathrm{mt}_\mathrm{p}$ is also evident, but only clearly discernible for the first six bins ($\Delta_1 - \Delta_6$).       

\section{Summary and Discussion}
Whether it be through direct observations, forward modelling, or inversions, in order to be able to confidently interpret helioseismic observations and inferences made in regions of strong magnetic field, the actual physical effects of near-surface magnetic fields on ray propagation must be better understood and taken into account when analyzing or modelling active region sub-photospheres. Our approach here is akin to forward modelling of rays, but in a simulated sunspot atmosphere based on IVM surface magnetic-field profiles with a peak field strength of 3~kG and an external field-free Model S atmosphere used as the background or unperturbed medium. The main aim of these simulations was to isolate and understand the effects of the wave-speed inhomogeneities produced by the magnetic field from those thought to be produced from thermal or flow perturbations. 

The magneto-acoustic rays were propagated across the sunspot radius for a range of depths to produce a skip distance geometry similar to centre-to-annulus cross-covariances used in time-distance helioseismology. The perturbations from the Model S atmosphere were calculated for each radial grid position and range of frequencies ($3.5-5$~mHz), then binned into 11 different skip-distance geometries of increasing size. A separate, yet similar, set of simulations was then produced to isolate the role played by thermal variations inside the sunspot atmosphere on the ray skip-distance and travel-time perturbation profiles. This was achieved by having the magnetic field switched off in the sunspot model, thus essentially maintaining a modified sound-speed structure, but with no calculations of the Alfv\'en speed. 

These \textit{artificial} skip-distance and travel-time perturbation profiles, which directly account for the effects near-surface magnetic fields and thermal variations separately, have provided us with a number of very distinct and interesting observations:
\begin{enumerate}
  \item The sunspot magnetic field has a clear and distinct ``dual effect'' on helioseismic rays -- increasing their skip distances, while at the same time, shortening their travel time (compared to similar rays in a Model S atmosphere). Higher frequency rays propagated within the magnetic field also tend to undergo a more substantial speed up than their non-magnetic counterparts.
	\item There is a clear and significant frequency dependence of both ray skip-distance and travel-time perturbations across the simulated sunspot atmosphere. This frequency dependence of perturbations was prevalent for all skip-distance bins, but particularly so for shallow rays, which sample the near-surface layers of the sunspot. 
	\item The negative sign of travel-time shifts, along with the general pattern and magnitude of these perturbations (\textit{i.e.} tending to increase with increasing magnetic-field strength and inclination) points to more evidence of the significant role played by the sunspot magnetic field. Rays with shorter skip distances were seen to experience greater perturbations as a result of spending a considerable proportion of their journey within the confines of the magnetic field. 
	\item With the magnetic field switched off, the simulated travel-time perturbation profiles changed sign for all bins (\textit{i.e.} only positive perturbations were observed across the sunspot radius, meaning that rays in the thermal model are actually slower than their Model S counterparts), and the magnitude of these perturbations appeared to be significantly smaller in magnitude ($300$--$400$\% at times) than when the magnetic field is included in the model. This was particularly evident for the bins that sample rays in the near-surface layers, whereas bins of larger skip distances produce slightly larger perturbations than the magnetic model. Frequency dependence of travel-time perturbations were also observed, but only for half of the bins. A majority of bins sampling larger skip distances did not exhibit this behaviour. 
\end{enumerate}

These observation as a whole tend to suggest that active-region magnetic fields play a direct and significant role in sunspot seismology, and it is the interaction of the near-surface magnetic field with solar oscillations, rather than purely thermal (or sound-speed) perturbations, that is the major cause of observed travel-time perturbations in sunspots. (We note here that we are only commenting on the interpretation of time-distance results in terms of \textit{thermal/sound-speed} perturbations, and not, for example, in terms of \textit{wave-speed} perturbations).

The frequency dependence of these perturbations is one of the strongest indications that the magnetic field is a significant contributor to the travel-time shifts. When isolating the thermal component of $\updelta\tau_\mathrm{p}$ we did observe some frequency dependence in a limited number of bins/skip distance geometries, certainly not to the extent that we saw when the magnetic field was included. Of course in the absence of any perturbations, rays propagated at different frequencies will naturally have slightly different upper turning points, this could certainly explain a part of a frequency dependence, but this effect combined with the (negative) sign and magnitude of the simulated $\updelta\tau_\mathrm{p}$ profiles, along with the relatively small (positive) thermal component extracted from the perturbations, makes it very difficult for one to argue that what we are seeing in these travel-time perturbation profiles is a result of a sub-surface flow or sound-speed perturbation, as has been traditionally interpreted in time-distance literature.

Instead, these observations indicate that strong near-surface magnetic fields may be seriously altering the magnitude and lateral extent of sound-speed inversions made by time-distance helioseismology. This is because standard time-distance observations (\textit{e.g.} \inlinecite{cbk06}, see Section 4.2) show $\updelta\tau^\mathrm{m}_\mathrm{p}$ maps derived from the averaged cross-correlations shifting from positive values for the first couple of bins (usually $\Delta_1 - \Delta_3$), to negative ones for the remainder of the bins. Traditionally, positive perturbations result in regions of decreased sound speed in inversions, while negative perturbations result in regions of enhanced sound speed. But we have clearly seen from our forward modelling that the inclusion of the magnetic field in the near surface layers consistently results in negative values for all bins of $\updelta\tau^\mathrm{m}_\mathrm{p}$. This implies that any inversion of time-distance data that does not account for surface magnetic field effects will be significantly contaminated in the shallower layers of the sunspot (\textit{i.e.} down to a depth of a few Mm below the surface), in strong agreement with the conclusions of \inlinecite{sebraj}. Hence it is almost certain from these simulations that the two-structure sunspot sound speed profile, \textit{i.e.} region of decreased sound speed immediately below the sunspot (corresponding to positive $\updelta\tau^\mathrm{m}_\mathrm{p}$), is most likely an artifact due to surface effects, instead of thermal perturbations. Deeper sound speed profiles do not appear to be affected as much, given the sign and magnitude of the simulated $\updelta\tau^\mathrm{m}_\mathrm{p}$ for the larger bins are comparable to actual time-distance calculations, as expected, given the flux tube becomes gas-pressure dominated at such depths.   

Of course, we must bear in mind that some of our assumptions outlined earlier (\textit{e.g.} 2D treatment of rays, our choice of ray cutoff height in the atmosphere, the fact that we are not directly accounting for mode conversion, even the form of the surface magnetic field and background model in general \textit{etc.}), can certainly alter our results quantitatively in one manner or another. Indeed it would certainly be interesting and worthwhile to conduct a full 3D simulation (\textit{i.e} vary the shooting angle $\beta$ around the sunspot) and also test the ray propagation code with other sunspot and quiet-Sun models in the future. But in any case, it would be surprising, given the self-consistency of our current results, if our qualitative conclusions were changed as a result.  

%% Acknowledgements
\begin{acks}
The authors are very grateful to Hannah Schunker for providing the IVM observations of AR 9026, and also to S\'ebastien Couvidat for providing us with the travel-time maps of AR 8243. We also thank the anonymous referee for useful comments that helped improve this paper. 
\end{acks}

\end{article} 
\end{document}